\documentclass[12pt]{article}

\usepackage{amssymb, amsmath, amsthm, amsfonts, graphicx, color, bm, float} 
\usepackage[margin=1in]{geometry}
\usepackage{hyperref}
\usepackage{chicago}
\usepackage{setspace} 
\usepackage{graphicx, subfig, epstopdf, enumitem}
\usepackage{caption}

\usepackage{algorithm, algpseudocode}

\newfloat{algorithm}{htbp}{loa}
\floatname{algorithm}{Algorithm}

\title{High-dimensional ABC}

\author{D. J. Nott\footnote{Department of Statistics and Applied Probability, National University of Singapore},\:\: V. M.-H. Ong$^{*}$,\:\: Y. Fan\footnote{School of Mathematics and Statistics, University of New South Wales, Sydney.}\:\:  and S. A. Sisson$^{\dagger}$}

\begin{document}

\maketitle

\section{Introduction}
\label{intro}

Other chapters in this volume have discussed the curse of dimensionality that is inherent to most standard ABC methods.  For a $p$-dimensional parameter of interest $\theta=(\theta_1,\dots,\theta_p)^\top$, ABC implementations make use of a summary statistic $s=S(y)$ for data $y\in {\cal Y}$ of dimension $q$, where typically $q\geq p$. When either $\theta$ or $s$ is high dimensional,  standard ABC methods have difficulty in producing  simulated summary data that are acceptably close to the observed summary $s_{obs}=S(y_{obs})$, for observed data $y_{obs}$. This means that standard ABC methods have limited applicability in high dimensional problems.

More precisely, write $\pi(\theta)$ for the prior, 
$p(y|\theta)$ for the data model, $p(y_{obs}|\theta)$ for the likelihood function and $\pi(\theta|y_{obs})\propto p(y_{obs}|\theta)\pi(\theta)$ for the intractable posterior distribution.
Standard ABC methods based on $S(y)$ typically approximate the posterior as
$\pi(\theta|y_{obs})\approx \pi_{ABC,h}(\theta|s_{obs})$, where  
\begin{align}
 \pi_{ABC,h}(\theta|s_{obs})& \propto \int K_h(\|s-s_{obs}\|)p(s|\theta)\pi(\theta)\,ds, 
 \label{kapprox}
\end{align}
and where $K_h(\|u\|)$ is a kernel weighting function with bandwidth $h\geq 0$.  
A Monte Carlo 
approximation of (\ref{kapprox}) involves a kernel density estimation of the 
intractable likelihood based on $\|s-s_{obs}\|$, the distance between simulated and observed summary statistics. As a result,  the quality of the approximation decreases rapidly as the dimension of the summary statistic $q$ increases, as the distance between $s$ and $s_{obs}$ necessarily increases with their dimension, even setting aside the approximations involved in the choice of an informative $S(y)$.  

Several authors (e.g. \shortciteNP{blum10}, \shortciteNP{barber+vw15}) have given results which illuminate the way that the dimension of the summary
statistic  $q$ impacts the performance of standard ABC methods.  For example, \citeN{blum10} obtains the result that the minimal mean squared error
of certain kernel ABC density estimators is of the order of $N^{-4/(q+5)}$, where $N$ is the number of Monte Carlo samples 
in the kernel approximation.  \shortciteN{barber+vw15} consider a simple rejection ABC algorithm where the kernel $K_h$ is uniform,
and obtain a similar result concerned with optimal estimation of posterior expectations.  
\shortciteN{biau+cg15} extend the analysis of \citeN{blum10} using
a nearest neighbour perspective, which accounts for the common ABC practice of choosing $h$ adaptively based on a large pool of samples (e.g. \shortciteNP{blum+nps13}).

Regression adjustments  (e.g. \shortciteNP{blum10,beaumont+zb02,blum+f10,blum+nps13}) are extremely valuable in practice for extending the applicability
of ABC approximations to higher dimensions, since the regression model has some ability to compensate for the mismatch between the simulated
summary statistics $s$ and the observed value $s_{obs}$.   \index{Regression adjustment}
However, except when the true relationship between $\theta$ and $s$ is known precisely (allowing for a perfect adjustment), these approaches may only extend ABC applicability to moderately higher dimensions. For example, \shortciteN{nott+fms14} demonstrated a rough doubling of the number of acceptably estimated parameters for a fixed computational cost when using regression adjustment compared to just rejection sampling, for a simple toy model.
Nonparametric regression approaches are also subject
to the curse of dimensionality, and the results of \citeN{blum10}
also apply to certain density estimators which include nonparametric regression adjustments.  Nevertheless, it has been
observed that these theoretical results may be overly pessimistic in practice for some problems.  
See  \citeN{li+f16} for some recent progress on theoretical aspects of regression adjustment for uncertainty quantification.

This chapter considers the question of whether it may be possible to conduct reliable ABC-based inference for high-dimensional models, or when the number of summary statistics $q\geq p$ is large. 
As a general principle, any methods that improve the efficiency of existing ABC techniques, such as more efficient Monte Carlo sampling algorithms, will as a result help extend ABC methods to higher dimensions, simply because they permit a greater inferential accuracy (measured by an effectively lower kernel bandwidth $h$) for the same computational overheads. However there is a limit to the extent to which these improvements can produce substantial high-dimensional gains, as ultimately the bottleneck is determined by the $\|s-s_{obs}\|$ term within the kernel $K_h$ embedded as part of the approximate posterior $\pi_{ABC,h}(\theta|s_{obs})$.

Instead, we examine ways in which the reliance on the $q-$dimensional comparison $\|s-s_{obs}\|$ can be reduced. One technique for achieving this is
by 
estimating low-dimensional marginal posterior distributions for subsets of $\theta$ and then reconstructing an estimate
of the joint posterior distribution from these. This approach takes advantage of the fact that the marginal posterior distribution 
$\pi_{ABC,h}(\theta^{(1)}|s_{obs})=\int \pi_{ABC,h}(\theta|s_{obs})d\theta^{(2)}$ for some partition of the parameter vector $\theta=(\theta^{(1)^\top},\theta^{(2)^\top})^\top$
can be much more accurately approximated using ABC directly as $\pi_{ABC,h}(\theta^{(1)}|s^{(1)}_{obs})$, since the corresponding necessary set of summary statistics $s^{(1)}\subset s$ would be a lower 
dimensional vector compared with the summary statistics $s$ required to estimate the full joint distribution $\pi_{ABC,h}(\theta|s_{obs})$. 
The same idea can also be implemented when
approximating the likelihood function, where it is the sampling distribution of the summary statistics $p(s|\theta)$ that is approximated
based on low-dimensional estimates for subsets of $s$. 

The above techniques are applicable for general ABC inference problems without any particular
exploitable model structure, and are the primary focus of this chapter.  For models with a known exploitable structure it may be possible to achieve better results (e.g. \shortciteNP{barthelme+c14,white+kp15,bazin+db10,tran+nk16,ong+ntsd16}), and we also discuss these briefly.

\section{Direct ABC approximation of the posterior}

In this section we consider direct approximation of the posterior distribution $\pi(\theta|s_{obs})$ given the observed summary statistics $s_{obs}$.  
We first describe the {\em marginal adjustment} approach of \shortciteN{nott+fms14}, in which the standard ABC approximation of the joint posterior distribution is improved by replacing its univariate margins with more precisely estimated marginal approximations.
These more precise marginal distributions are obtained by implementing standard ABC methods to construct
each univariate marginal posterior separately, for which only low-dimensional summary statistics are required. These univariate marginal posteriors 
then replace the margins in the original approximate joint posterior sample, via an appropriate replacement of order statistics.

While the marginal adjustment can work well
we show an instructive toy example where this strategy fails to adequately estimate 
the posterior dependence structure. We subsequently discuss the Gaussian copula ABC approach of  \shortcite{li+nfs15}, which extends the marginal adjustment to improve estimation of all
pairwise dependences of the joint posterior, in combination
with the marginal estimates, by use of a meta-Gaussian distribution \shortcite{fang+fk02}. 
These ideas are illustrated by several examples.

\subsection{The marginal adjustment strategy}\label{mas}

The marginal adjustment method of \shortciteN{nott+fms14} is motivated by the following observation.  \index{Marginal adjustment} Suppose we wish to estimate accurately 
the univariate marginal posterior distribution $\pi(\theta_j|s_{obs})$ of the parameter $\theta_j$. If we can find a summary statistic, say $s^{(j)}\subset s$, that is nearly marginally sufficient for $\theta_j$ in the data model $p(y|\theta_j)$, then $\pi(\theta_j|s_{obs}) \approx \pi(\theta_j|s^{(j)}_{obs}) $ and
this summary statistic can be used to obtain marginal ABC posterior inferences about $\theta_j$.    Because $\theta_j$ is univariate, the summary statistic $s^{(j)}$ can be low-dimensional.  

Accordingly, the marginal ABC model takes the form
$$\begin{array}{ll}
\pi_{ABC, h}(\theta_j|s^{(j)}_{obs}) &\propto \int K_h(\|s^{(j)}-s^{(j)}_{obs} \|)p(s^{(j)}|\theta_j)  \pi(\theta_j) ds^{(j)}\\
&=\int\int K_h(\|s^{(j)}-s^{(j)}_{obs} \|)   p(s|\theta) \pi(\theta_{-j}|\theta_j)  \pi(\theta_j) d\theta_{-j}  ds 
\end{array}  $$
where $\theta_{-j}$ denotes the elements of $\theta$ excluding $\theta_j$, and $\pi(\theta_{-j}|\theta_j)$ denotes the conditional prior
of $\theta_{-j}$ given $\theta_j$.

The idea of \shortciteN{nott+fms14} is to exploit the observation that marginal posterior inferences are much easier in the ABC framework as they only involve a 
lower dimensional  subset of summary statistics, $s^{(j)}\subset s$.
A sample from the joint ABC posterior $\pi_{ABC, h}(\theta|s_{obs})$ is first obtained, and then this joint sample is adjusted so that it's marginal distributions match
those estimated from the lower-dimensional ABC analyses, $\pi_{ABC, h}(\theta_j|s^{(j)}_{obs})$.

Write $s=(s_1,\dots,s_q)^\top$ for the summary statistics used to approximate the joint posterior $\pi_{ABC, h}(\theta_j|s_{obs})$, and $s^{(j)}=(s_1^{(j)},\dots,s^{(j)}_{q_j})^\top$ for the summary statistics used to approximate the marginal posterior distribution of $\theta_j$, $\pi_{ABC, h}(\theta_j|s^{(j)}_{obs})$.
The marginal adjustment algorithm is then implemented as follows: 
\begin{enumerate}
\item Using standard ABC methods (including regression adjustments) obtain an approximate sample from the joint posterior distribution $\pi(\theta|s_{obs})$, $\theta^{J1},\dots,\theta^{Jr}$ say, based on the full summary statistic $s$.  
\item Using standard ABC methods, for each $j=1,\dots,p$, obtain an approximate sample from the univariate marginal distribution $\pi(\theta_j|s^{(j)}_{obs})$, $\theta_j^{M1},\dots,\theta_j^{Mr'}$ say, based on the lower-dimensional summary statistic $s^{(j)}$.
\item Write $\theta_{j}^M(k)$ for the $k$-th order statistic of the (marginally estimated) sample $\theta_{j}^{M1},\dots,\theta_j^{Mr'}$ and 
$\theta_{j}^J(k)$ for the $k$-th order statistic of the (jointly estimated marginal) sample $\theta_j^{J1},\dots,\theta_j^{Jr}$. Also write 
$R(j,k)$ for the rank of $\theta_j^{Jk}$ within the sample $\theta_j^{J1},\dots,\theta_j^{Jr}$.  Define
$$\theta^{Ak}=(\theta^M_1(R(1,k)),\dots,\theta^M_p(R(p,k)))^\top.$$
Then $\theta^{Ak}$, $k=1,\dots,r,$ is a marginally adjusted approximate sample from $\pi(\theta|s_{obs})$.  
\end{enumerate}
It is worth stating in words what is achieved by step 3 above.  The samples $\theta^{Ak}$, $k=1,\dots,r$ are the same as
$\theta^{Jk}$, except that componentwise  the order statistics $\theta_{j}^J(k)$ have been replaced by the 
corresponding order statistics $\theta_{j}^M(k)$.  If we were to convert the samples $\theta^{Ak}$ and
$\theta^{Jk}$ to ranks componentwise they would be exactly the same, and so the dependence structure in the original
samples $\theta^{Jk}$ is preserved in $\theta^{Ak}$ in this sense.  However, the estimated marginal distribution
in $\theta^{Ak}$ for $\theta_j$ is simply the estimated marginal distribution obtained from the samples $\theta_j^{M1},\dots,\theta_j^{Mr'}$, 
so that the adjusted samples $\theta^{Ak}$ give the more precisely estimated marginal distributions from the low-dimensional analyses of step 2, while
preserving the dependence structure from the joint samples of step 1.  

While it is true that the dependence structure obtained at step 1 may not
be well estimated due to standard ABC curse-of-dimensionality arguments, it is also the case that the marginal adjustment improves the estimation of the marginal posterior distributions.
These ideas are illustrated in the following example.

\subsection{A toy example}\label{toy}

\begin{figure}
\centering
\includegraphics[scale=0.44]{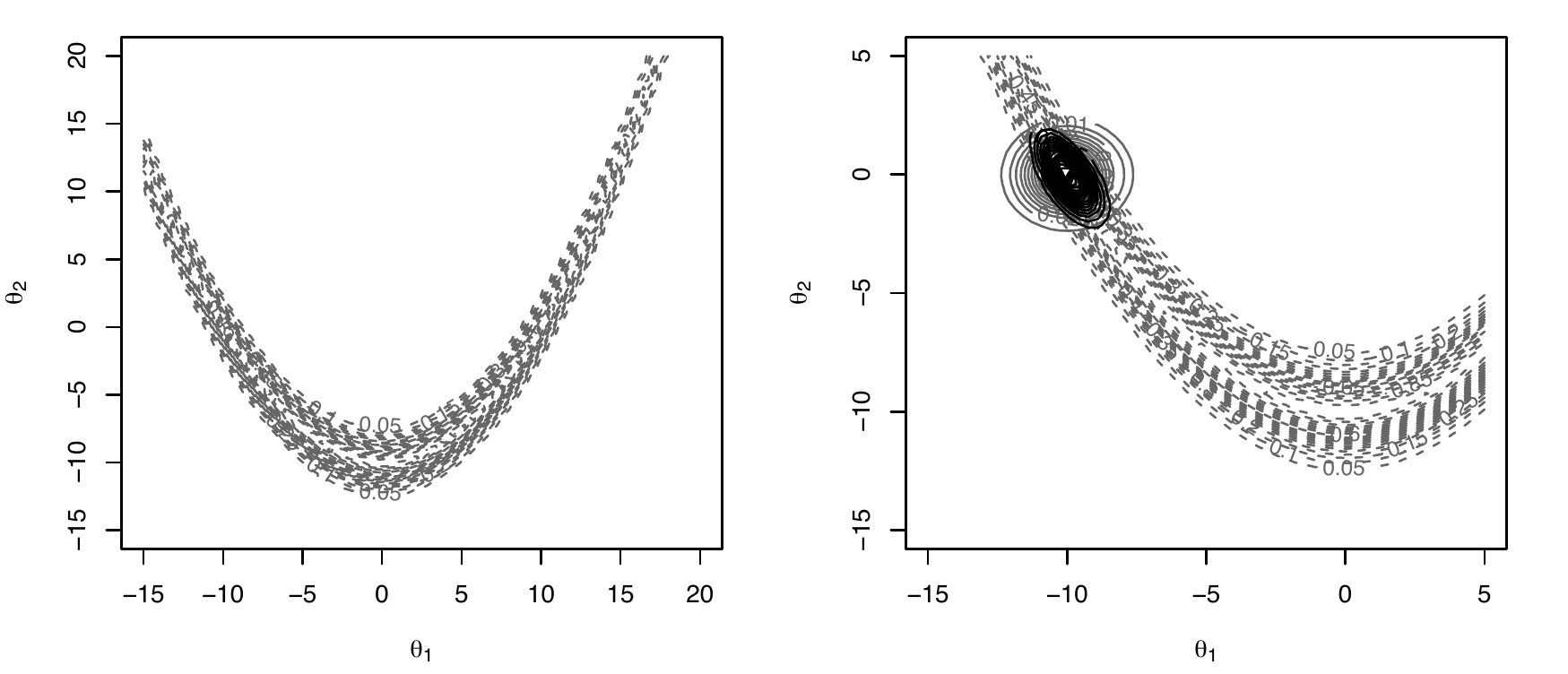}
\includegraphics[scale=0.44]{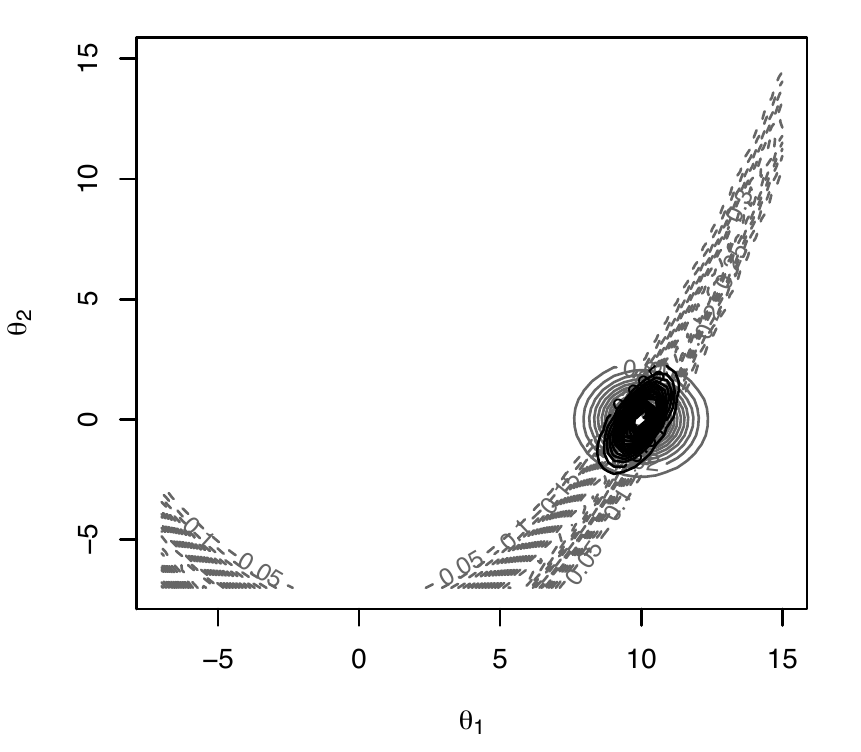}
\caption{\small Contour plots of twisted normal prior distribution $\pi(\theta)$ (grey dashed lines), likelihood (solid grey) and posterior (solid black) for $p=2$. 
Middle and right panels illustrate the case when $y_{obs} = (-10,0)^\top$ and $y_{obs} = (10,0)^\top$ respectively. 
}\label{fig:contour prior}
\end{figure}

Following \shortciteN{li+nfs15} we let  the data $y=(y_1,\dots,y_p)^\top$, $p\geq 2$ follow a $N(\theta,I_p)$ distribution where $\theta=(\theta_1,\dots,\theta_p)^\top$ is the parameter of interest and $I_p$ denotes the $p\times p$ identity matrix. The prior $\pi(\theta)$ is specified as the twisted normal form \shortcite{haario+st99}
$$\pi(\theta)\propto \exp\left(-\frac{\theta_1^2}{200}-\frac{(\theta_2-b\theta_1^2+100 b)^2}{2}-\sum_{j=3}^p \theta_j^2\right)$$
where we set $b=0.1$, and if $p=2$ the $\sum_{j=3}^p \theta_j^2$ term is omitted.  
A contour plot of $\pi(\theta)$ for $p=2$ is shown in Figure \ref{fig:contour prior}.  
This is an interesting example because the likelihood only provides location information
about $\theta$.  The dependence structure in the posterior comes mostly from the prior, and the assocation between $\theta_1$ and
$\theta_2$ changes direction in the left and right tails of the prior (Figure \ref{fig:contour prior}).  So the posterior dependence changes direction
depending on whether the likelihood locates the posterior in the left or right tail of the prior.  This feature makes it difficult for
standard regression adjustment methods, which merely translate and scale particles (e.g. generated from $(s,\theta)\sim p(s|\theta)\pi(\theta)$), to work in high-dimensions.

\begin{figure}[tb]
\centering
\includegraphics[scale=0.55]{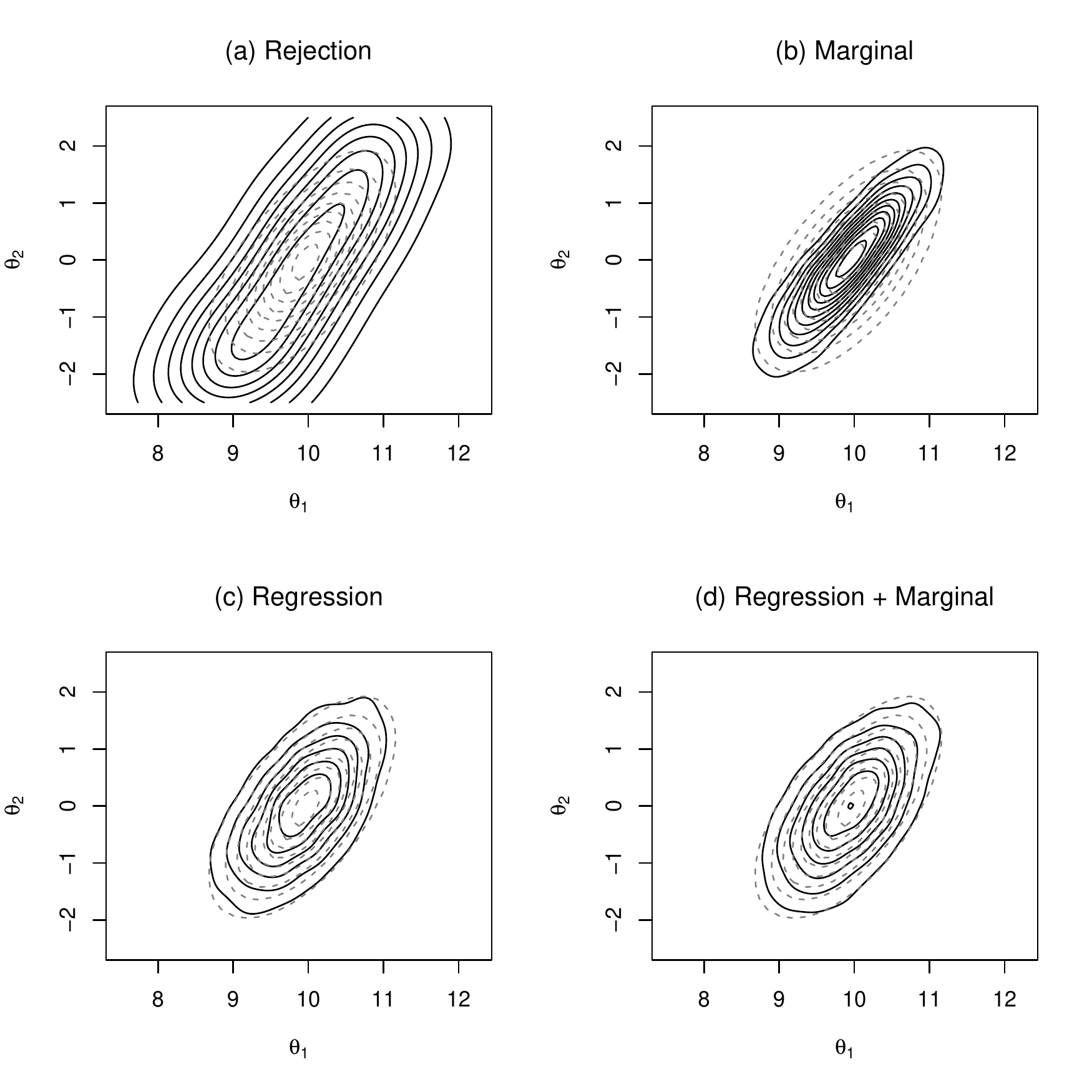}
\caption{\small Contour plots of the $(\theta_1,\theta_2)$ margin of various ABC posterior approximations for the $p=5$ dimensional model $\pi(\theta|s_{obs})$ are represented by the black lines. True contours for the bivariate margins are represented by the grey-dashed lines. The different ABC approximations approaches are (a) rejection sampling, (b) rejection sampling with marginal adjustment, (c) rejection sampling with regression adjustment and (d) rejection sampling with regression and marginal adjustment. \label{p5}}
\end{figure}

\begin{figure}[tb]
\centering
\includegraphics[scale=0.55]{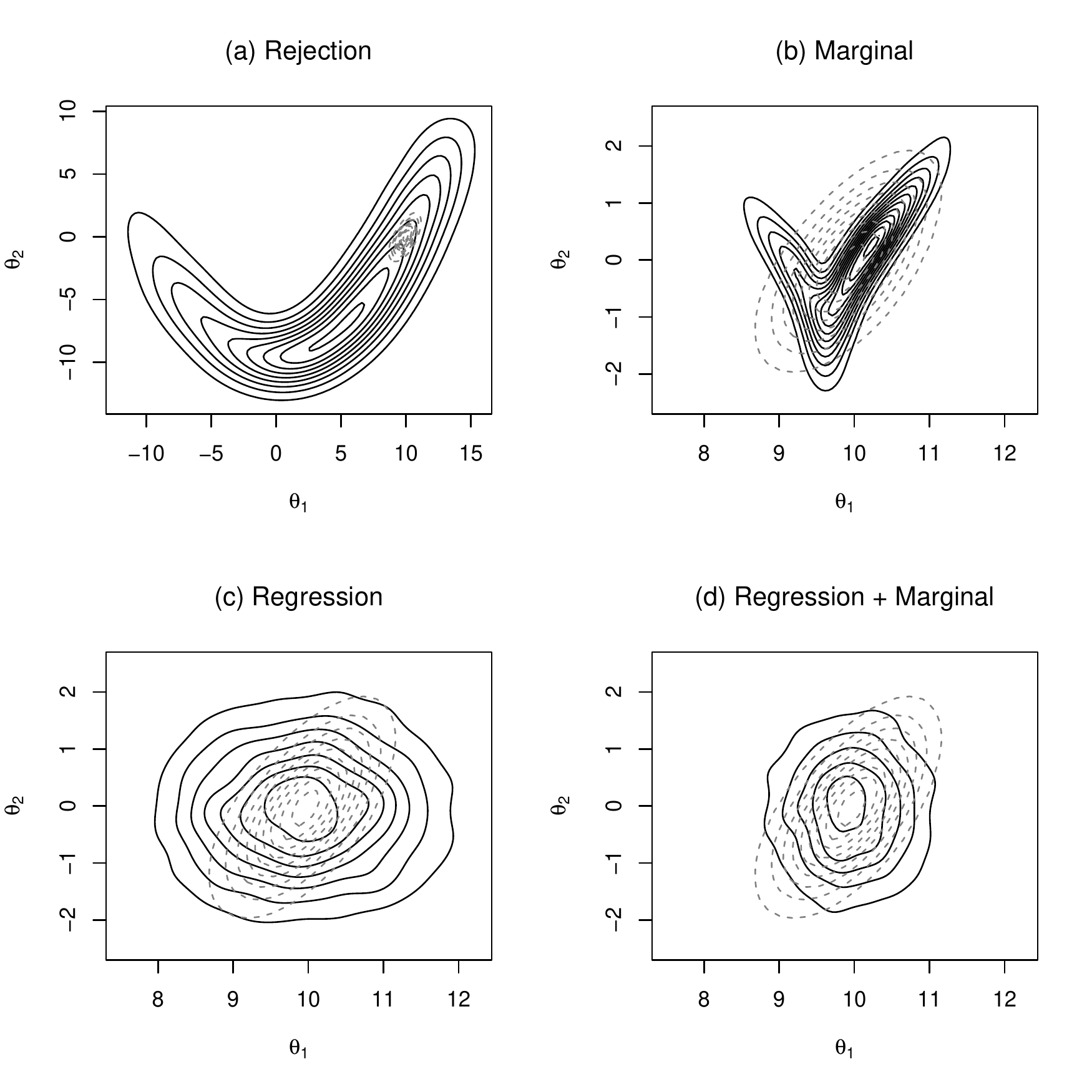}
\caption{\small Contour plots of the $(\theta_1,\theta_2)$ margin of various ABC posterior approximations for the $p=50$ dimensional model $\pi(\theta|s_{obs})$ are represented by the black lines. True contours for the bivariate margins are represented by the grey-dashed lines. The different ABC approximations approaches are (a) rejection sampling, (b) rejection sampling with marginal adjustment, (c) rejection sampling with regression adjustment and (d) rejection sampling with regression and marginal adjustment. \label{p50}}
\end{figure}

Figures \ref{p5} and \ref{p50} show what happens in an analysis of this example with $p=5$ and $p=50$ respectively. Four ABC approximation methods are considered with observed data $y_{obs} = (10,0,...,0)^\top$. The contour plots of the bivariate posterior estimates $\pi(\theta_1,\theta_2|s_{obs})$ are represented by solid lines while the contour plot of the true bivariate margin is represented by grey dashed lines. For both Figures, panel (a) shows the estimates obtained via standard rejection ABC while panels (b), (c) and (d) show estimates obtained after marginal, linear regression and both linear regression and marginal adjustment respectively. Note the regression adjustment step is performed after the rejection sampling stage, and before the marginal adjustment. 

For the case when $p=5$ (Figure \ref{p5}), rejection sampling alone captures the correlation between $\theta_1$ and $\theta_2$, but the univariate margins are too dispersed.  Performing a marginal adjustment following rejection sampling is not good enough, as it only corrects the margin to the right scale and is not able to recover dependence structure. On the other hand, rejection sampling with linear regression adjustment is able to give a good approximation to the true posterior. Performing a subsequent marginal adjustment (Figure \ref{p5}(d)) shows no further visual improvement. 

The example for $p=50$ (Figure \ref{p50}) shows both the strengths and limitations of the marginal and regression adjustment strategies. It is very clear that standard rejection ABC estimates do not seem to be using much of the information given by the likelihood, as the posterior estimate follows the shape of the prior distribution $\pi(\theta)$. Performing either regression or marginal adjustment centres the estimates on the right location but the shape of the contour plots for the adjustments are incorrect. Moreover, applying marginal adjustment after regression adjustment corrects the univariate margins well, but does not recover the dependence structure. In this example, all four approaches are not able to recover the dependence structure of the true bivariate posterior. It is worth noting in this example that for the normal case with $b=0$ the marginal adjustment approach works very well even in high dimensions. 

This example shows some of the limitations of the marginal adjustment strategy. 
One possible approach to improve the estimation of the dependence structure (not discussed by \shortciteNP{nott+fms14}) is to use the marginal adjustment on a reparameterised parameter vector $\theta^*$, where the margins of $\theta^*$ account for the dependence structure in $\theta$, while  $\theta^*_i$ and $\theta^*_j$, $i\neq j$ remain approximately independent. 
This approach would require some prior knowledge of the dependence structure.

Since the key idea of the marginal adjustment approach is to build up a more accurate approximation of the joint posterior from estimates of univariate marginal posterior distributions, it is natural to ask if it is possible to consider estimation of marginal posterior distributions of dimension larger than one and to use these
to help estimate the joint dependence structure of $\pi(\theta|s_{obs})$ more accurately.

\subsection{Gaussian copula ABC}\label{gca}

One way to implement this idea is the Gaussian copula ABC method of \shortciteN{li+nfs15}.  \index{Gaussian copula ABC} Suppose that
$\mathcal{C}(u)=P(U_1\leq u_1,\dots, U_p\leq u_p)$ is the distribution function of a random vector $U=(U_1,\dots,U_p)$ where the marginal distribution of each
$U_j\sim U(0,1)$ is uniform.  Then $\mathcal{C}(u)$ is called a copula.  Multivariate distributions can always be written in terms of a copula and their marginal distribution functions, which is an implication of Sklar's theorem \cite{sklar59}. This allows for a decoupling of the modelling of marginal distributions and the dependence structure of a multivariate distribution.  One useful type of copula derives from a multivariate Gaussian distribution.  
Suppose that $\eta\sim N(0,C)$ is a $p$-dimensional multivariate Gaussian random vector where $C$ is
a correlation matrix.  The distribution of $U=(\Phi(\eta_1),\dots,\Phi(\eta_p))^\top$ where $\Phi(\cdot)$ denotes the standard normal distribution function
is then a copula.  This kind of copula, called a Gaussian copula, characterises the dependence structure of a multivariate Gaussian distribution and it is parametrised
by the correlation matrix $C$.

Suppose now that we further transform $U$ as
$\gamma=(F_1^{-1}(U_1),\dots,F_p^{-1}(U_p))^\top$ where $F_1(\cdot),\dots,F_p(\cdot)$ are distribution functions with corresponding density 
functions $f_1(\cdot),\dots,f_p(\cdot)$. The components of $\gamma$ then have the marginal densities $f_1(\cdot),\dots,f_p(\cdot)$ respectively, 
and the dependence structure is being described by the Gaussian copula with correlation matrix $C$.  Clearly if the densities $f_j(\cdot)$, $j=1,\dots,p$ are
themselves univariate Gaussian then $\gamma$ is multivariate Gaussian.  A distribution constructed from a Gaussian copula and given marginal distributions 
is called {\it meta-Gaussian} \shortcite{fang+fk02} and its density function is
$$h(\gamma)=|C|^{-1/2}\exp\left(\frac{1}{2} z^\top(I-C^{-1})z \right)\prod_{j=1}^p f_j(\gamma_j)$$
where $z=(z_1,\dots,z_p)^\top$ and $z_j=\Phi^{-1}(F_j(\gamma_j))$.  

\shortciteN{li+nfs15} considered using a meta-Gaussian distribution to approximate the posterior distribution $\pi(\theta|s_{obs})$ in ABC.  It is easily seen that a meta-Gaussian distribution is 
determined by its bivariate marginal distributions, so that if we are prepared to accept a meta-Gaussian approximation to the joint posterior distribution
in a Bayesian setting, then it can be constructed based on bivariate posterior marginal estimates.   
Asymptotically the posterior will tend to be Gaussian, but a meta-Gaussian approximation may work well even when we are far from
this situation since it  allows for flexible estimation of the marginal distributions.  As with the marginal adjustment, since the bivariate marginal posterior distributions can 
be estimated using low-dimensional summary statistics, this can help to circumvent the ABC curse of dimensionality in estimation of the joint posterior dependence
structure. 

As before, write $s^{(j)}$ for the statistics that are informative for ABC estimation of the univariate posterior marginal $\pi(\theta_j|s_{obs})$, and now write $s^{(i,j)}$ for the summary statistics informative
for ABC estimation of the bivariate posterior margin $\pi(\theta_i,\theta_j|s_{obs})$, $i\neq j$.  Construction of the Gaussian copula ABC approximation to the posterior $\pi(\theta|s_{obs})$ proceeds as follows:
\begin{enumerate}
\item\label{111} Using standard ABC methods (including regression adjustments), for each $j=1,\dots,p$, obtain an approximate sample from the univariate marginal distribution  $\pi(\theta_j|s^{(j)}_{obs})$,  $\theta_j^{U1},\dots,\theta_j^{Ur}$ say, based on the lower dimensional summary statistic $s^{(j)}$.  Use kernel density estimation to construct an approximation $\hat{g}_j(\theta_j)$ to $\pi(\theta_j|s^{(j)}_{obs})$.
\item\label{222} Using standard ABC methods, for $i=1,\dots,p-1$ and $j=i+1,\dots,p$, obtain an approximate sample
from the bivariate marginal distribution $\pi(\theta_i,\theta_j|s^{(i,j)}_{obs})$, $(\theta_i^{Bj1},\theta_j^{Bi1}),\dots,(\theta_i^{Bjr},\theta_j^{Bir})$ say, based on the low-dimensional summary statistics
$s^{(i,j)}$.

\item Write $R(i,j,k)$ as the rank of $\theta_i^{Bjk}$ within the sample $\theta_i^{Bj1},\dots,\theta_i^{Bjr}$.  With this notation
$R(j,i,k)$, $j>i$, is the rank of
$\theta_j^{Bik}$ within the sample $\theta_j^{Bi1},\dots,\theta_j^{Bir}$.  Estimate $C_{ij}$ by $\hat{C}_{ij}$,  the sample correlation between the vectors
$$\left(\Phi^{-1}\left(\frac{R(i,j,1)}{r+1}\right),\Phi^{-1}\left(\frac{R(i,j,2)}{r+1}\right),\dots,\Phi^{-1}\left(\frac{R(i,j,r)}{r+1}\right)\right)^\top$$
and
$$\left(\Phi^{-1}\left(\frac{R(j,i,1)}{r+1}\right),\Phi^{-1}\left(\frac{R(j,i,2)}{r+1}\right),\dots,\Phi^{-1}\left(\frac{R(j,i,r)}{r+1}\right)\right)^\top.$$
\item Construct the Gaussian copula ABC approximation of $\pi(\theta|s_{obs})$ as  the meta-Gaussian distribution with marginal distributions $\hat{g}_j(\theta_j)$, $j=1,\dots,p$ 
(step \ref{111}), and Gaussian
copula correlation matrix $\hat{C}=[\hat{C}_{ij}]_{i,j=1,\dots,p}$ where $\hat{C}_{ij}$, $j>i$, is as in step \ref{222}, $\hat{C}_{ji}=\hat{C}_{ij}$ and $\hat{C}_{ii}=1$.  
\end{enumerate}
While the estimated correlation matrix $\hat{C}$ can fail to be positive definite using this procedure (although this did not occur in our analyses), methods to adjust this can be easily implemented e.g. \shortcite{loland+hhf13}. Note that by using
the approximate posterior sample from $\pi(\theta_i,\theta_j|s^{(i,j)}_{obs})$ from step 2 and the fitted (bivariate) copula model for the pair, it is possible to
investigate whether the Gaussian copula dependence structure at least represents the true bivariate posterior dependence structure well (though not the full multivariate dependence structure).  This can be supplemented
by  application specific goodness of fit checking of posterior predictive densities based on the joint copula approximation.

\begin{figure}[tb]
\centerline{\includegraphics[scale=0.25]{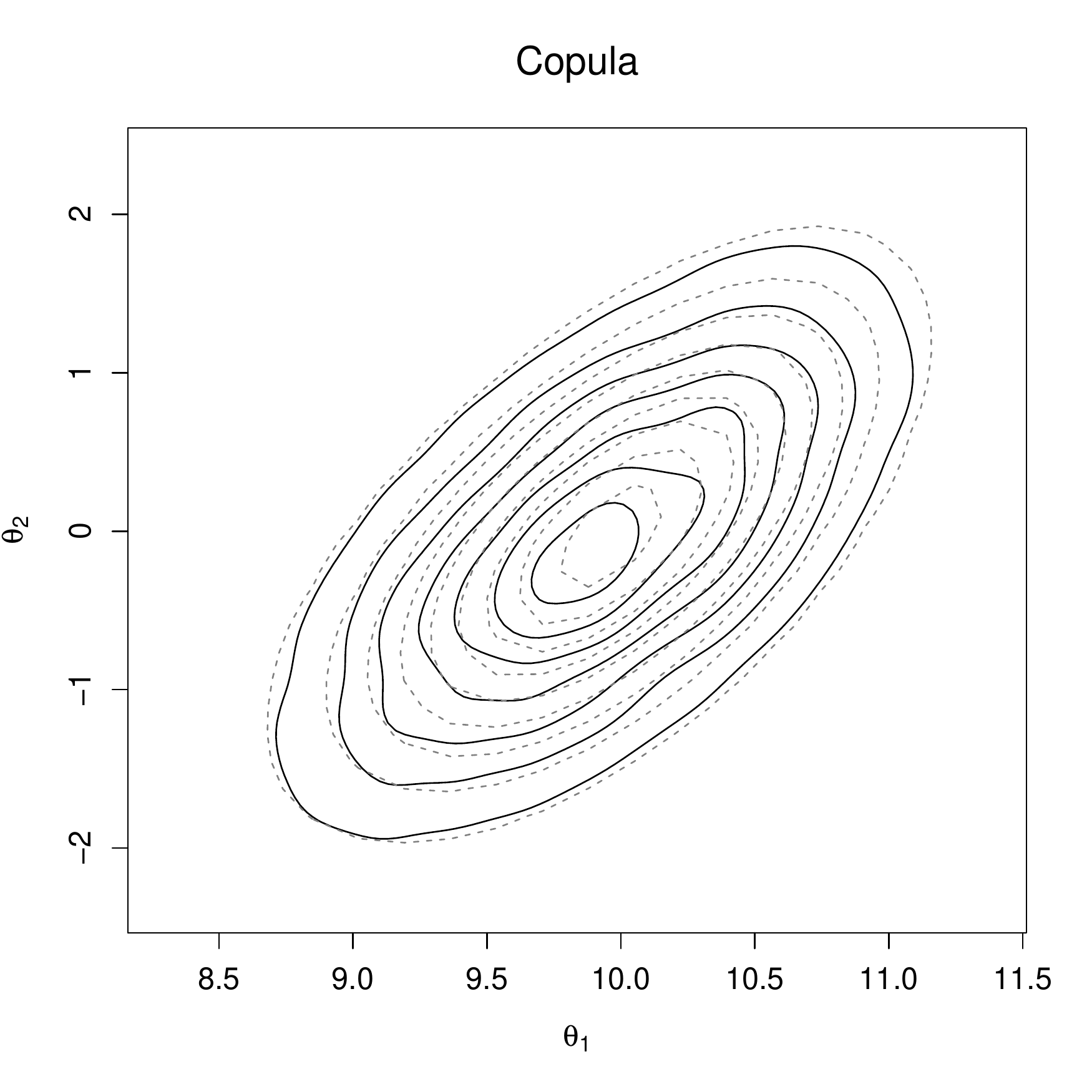}}
\caption{\small Contour plots of the $(\theta_1, \theta_2)$ margin of the Gaussian copula ABC posterior approximation of the $p=50$ dimensional model $\pi(\theta|s_{obs})$ (black lines). The true contours of $\pi(\theta_1,\theta_2|s_{obs})$ are represented by grey dashed lines.  \label{Copula}}
\end{figure}

In the twisted normal toy example of Section \ref{toy}, the copula strategy can succeed where the marginal adjustment strategy alone fails. 
Similar to Figure \ref{p50}, Figure \ref{Copula} illustrates both the bivariate estimates of $\pi(\theta_1,\theta_2|s_{obs})$  based on the Gaussian copula ABC approximation (black solid lines) and the true margins (grey dashed lines) for the $p=50$ dimensional model. From the contour plots, the ABC copula approximation is able to produce estimates largely similar to the true bivariate margins, in stark contrast to the marginal adjustment alone in Figure \ref{p50}. Thus, in this example where standard ABC sampling with regression and/or marginal adjustment fails, the copula strategy succeeds. 

In order to investigate the performance of each ABC posterior estimation method more precisely, we follow \shortciteN{li+nfs15} and vary the dimension of the model,  $p$, from $2$ to $250$. Table \ref{KL_Copula} shows the mean estimated Kullback-Leibler (KL) divergence between the true bivariate margin $\pi (\theta_1,\theta_2 | s_{obs})$ and the bivariate margin of the full ABC posterior approximation based on 100 replicate approximations, for all five approaches. 

Observe that when the dimension $p$ increases the performance of the standard rejection ABC approach deteriorates. Adopting any of the adjustment strategies improves the overall performance but the estimated KL divergences still increase with dimension $p$ up to fixed limits). This suggests that if accurate estimation of the posterior dependence structure is important, then regression and marginal adjustment strategies alone may be limited to low dimensional models. From Table \ref{KL_Copula} it is clear that Gaussian copula ABC outperforms all other methods in terms of KL divergence and its performance does not deteriorate with increasing dimension, $p$. This is not surprising as the Gaussian copula ABC approximation  is constructed from bivariate estimates of $\pi (\theta_1,\theta_2 | s_{obs})$, and is therefore able to capture the dependence structure of all bivariate pairs of the full posterior distribution $\pi(\theta|s_{obs})$.

\begin{table}
{\footnotesize
\begin{tabular}{l|lllll}
$p$ & Rejection & Marginal & Regression & Regression then  & Copula ABC \\
&only &&& Marginal & \\
 \hline
2   &   $0.058\: (<0.001)$    &  $0.040\: (<0.001)$ & $0.043\: (<0.001)$ & $0.035\:(<0.001)$ & $0.039\:(<0.001)$ \\
5   &   $0.807\: (<0.001)$    &  $0.053\: (0.001)$ & $0.613\: (0.002)$ & $0.037\:(<0.001)$ & $0.040\:(<0.001)$ \\
10  &   $1.418\: (0.002)$     &   $0.100\: (0.001)$ & $1.078\: (0.002)$ & $0.061\:(0.001)$ & $0.040\:(<0.001)$\\
15 &    $1.912\: (0.002)$     &   $0.292\: (0.002)$ & $1.229\: (0.003)$ & $0.202\:(0.001)$ & $0.039\:(<0.001)$\\
20   &  $2.288\: (0.002)$     &  $0.450\: (0.001)$ & $1.280\: (0.003)$ & $0.292\:(0.001)$ & $0.039\:(<0.001)$\\
50  &   $3.036\: (0.003)$     &  $0.520\: (0.002)$ & $1.474\: (0.009)$ & $0.335\:(0.001)$ & $0.040\:(<0.001)$\\
100  &  $3.362\: (0.002)$     &  $0.524\: (0.002)$ & $1.619\: (0.013)$ & $0.341\:(0.001)$ & $0.039\:(<0.001)$\\
250 &   $3.663\: (0.003)$    &  $0.515\: (0.002)$ & $1.737\: (0.015)$ & $0.344\:(0.001)$ & $0.039\:(<0.001)$
\end{tabular}}
\vspace{3pt}
\caption{\small Estimated Kullback-Leibler divergence of the $(\theta_1,\theta_2)$ margin of various ABC posterior approximations to $\pi (\theta_1,\theta_2 | s^{(1,2)}_{obs})$. Numbers in parentheses represent standard errors of mean divergences over 100 replications. }\label{KL_Copula}
\end{table}

In the following  sections we implement Gaussian copula ABC for two real data analyses: an analysis of multivariate currency exchange data, and simultaneous estimation of multiple quantile regressions.

\subsection{A multivariate $g$-and-$k$ model for a foreign currency exchange data set}
\label{sec:toy example} \index{$g$-and-$k$ distribution}

The $g$-and-$k$ distribution (\shortciteNP{raynerm02}) is a flexible model for univariate data. It is typically specified through its quantile function
\begin{eqnarray}
	\label{gk}
	Q(p|A,B,g,k)=A+B\left[1+c\frac{1-\mbox{exp}\{-g z(p)\}}{1+\mbox{exp}\{-g z(p)\}}\right](1+z(p)^2)^k z(p),
\end{eqnarray}
where $A,B>0,g$ and $k>-0.5$ are parameters respectively controlling location, scale, skewness and kurtosis of the distribution. The parameter $c$ is conventionally
fixed at $0.8$ (resulting in $k>-0.5$), and $z(p)$ denotes the $p$-quantile of the standard normal distribution.  Many distributions can be recovered or well approximated for
appropriate values of $A,B,g$ and $k$ (such as the normal when $g=k=0$).  Despite its attractive properties as a model, inference using the $g$-and-$k$ distribution is challenging since the density, given by the derivative of the inverse of the quantile function, has no closed form.  However, since simulation from the model is trivial by 
transforming uniform variates on $[0,1]$ through the quantile function, an ABC implementation is one possible inferential approach.  This idea was first explored by 
\citeN{peters+s06} and \shortciteN{allinghamkm09}. Here we consider a multivariate extension of the model developed by \citeN{drovandi+p11}.  
This model has a univariate $g$-and-$k$ distribution for each margin, and the dependence structure is specified through a Gaussian
copula.  Note that this use of a Gaussian copula to describe the dependence structure in the data model (likelihood) is distinct from the use of a Gaussian
copula to approximate the dependence structure of the posterior distribution.  

Suppose that the data are $n$ independent multivariate realisations $y=(y^1,\dots,y^n)$ where $y^i=(y^i_{1},\dots,y^i_{q})^\top$.  
We assume that 
marginally each $y^i_j$ $i=1,\ldots,n$ follows a
$g$-and-$k$ distribution with parameters $(A_j,B_j,g_j,k_j)$, $j=1,\dots,q$.  Gaussian copula ABC approximates the joint distribution of  $y^i$ by a meta-Gaussian distribution, 
with  Gaussian copula correlation matrix $C$.  For a $q$-dimensional data model, there are $4q$ marginal parameters, and $q(q-1)/2$ 
distinct parameters in the correlation matrix, 
giving $p=q(q +7)/2$ parameters in total.  
We consider 
an analysis of log daily returns for $q=16$ currencies  (resulting in $p=184$ parameters) versus the Australian dollar for 1757 trading days covering the period 1st January 2007 to 31st 
December 2013 \cite{rba14}.

We adopt as a prior on $C$ the distribution obtained by sampling $V\sim \mbox{Wishart}(I_q,q)$,
 and then rescaling $V$ to be a valid correlation matrix with $1$'s on the diagonal.  The priors on $A$, $B$, $g$ and $k$
for each marginal are independent and uniform over the parameter support, although we adopted uniform distributions with ranges of $[-0.1,0.1]$, $[0,0.05]$, $[-1,1]$ and $[-0.2,0.5]$ for $A_j$, $B_j$, $g_j$ and $k_j$ to produce samples $(s,\theta)$ proportional to $p(s|\theta)\pi(\theta)$ but restricted to a region of high posterior density following an initial pilot analysis (see e.g. \citeNP{fearnhead+p12}).

Following the strategy of \shortciteN{li+nfs15}, the following summary statistics
were considered informative for each marginal parameter:  writing $L_{kj}$, $k=1,2,3$ for the quantiles and $O_{kj}$, $k=1,\dots,7$ for the octiles of $y_j^1,\ldots,y_j^n$,
the marginally informative summary statistics were chosen as $L_{2j}$ for $A_j$, $(L_{3j}-L_{1j},(E_{7j}-E_{5j}+E_{3j}-E_{1j})/(L_{3j}-L_{1j}))^\top$ for
$B_j$, $(L_{3j}+L_{1j}-2L_{2j})/(L_{3j}-L_{1j})$ for $g_j$, and $(E_{7j}-E_{5j}+E_{3j}-E_{1j})/(L_{3j}-L_{1j})$ for $k_j$.  These summary statistic
choices were guided by similar summary statistics in \citeN{drovandi+p11} and preliminary analyses to determine which sets of the distinct summaries 
were marginally informative for individual parameters.  For pairs of parameters the summary statistics for individual parameters were simply combined.  
For the correlation parameters in the Gaussian copula, we follow \citeN{drovandi+p11} and use the robust normal scores correlation coefficient for the marginal summary statistic.

\begin{figure}
\centering
\includegraphics[scale=0.45]{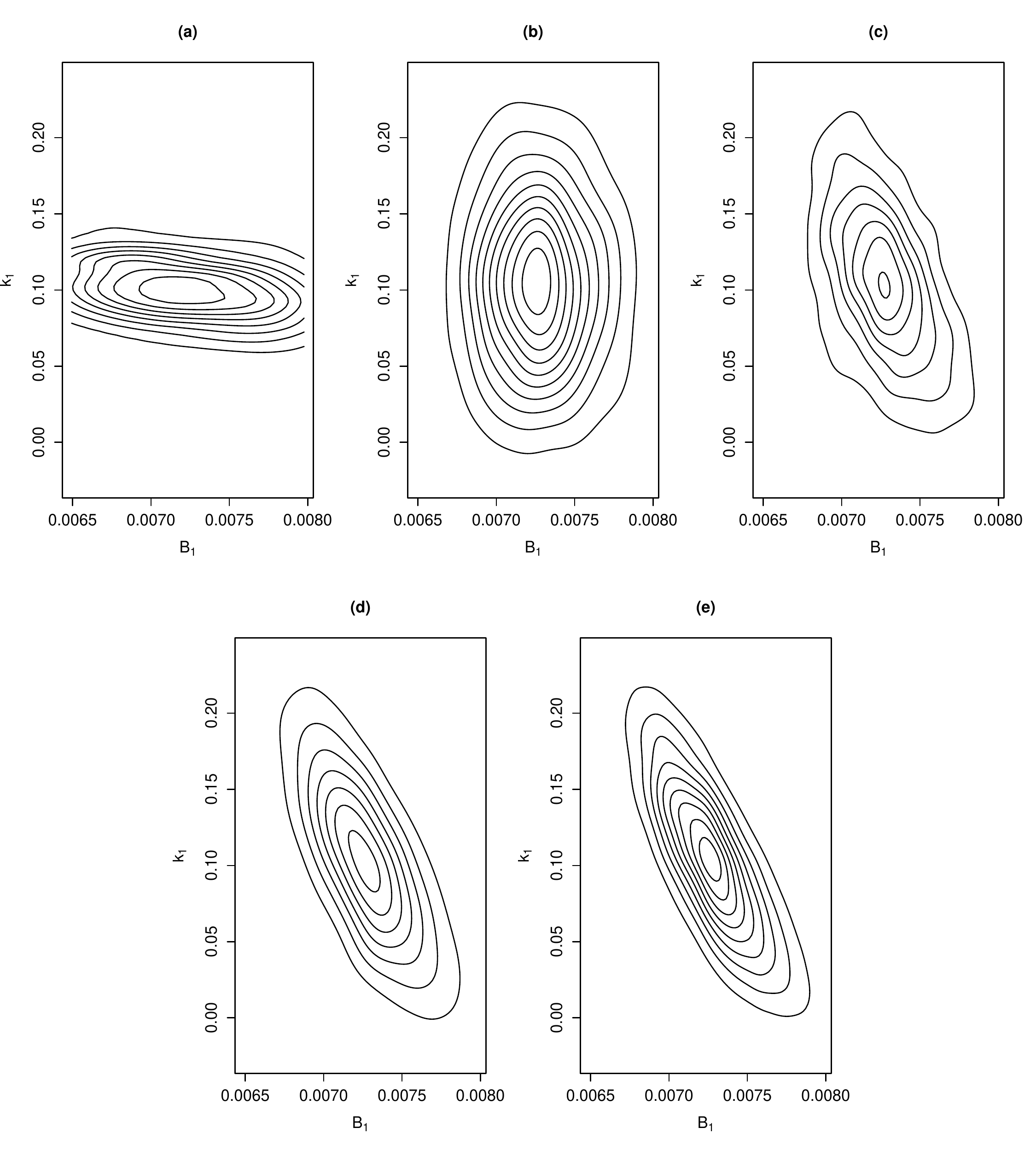}\vspace{-8mm}
\caption{\small Contour plots of the $(B_1,k_1)$ margin of rejection sampling based ABC posterior approximations to the multivariate $g$-and-$k$ model. 
Top panels show estimates using (a) regression adjustment, (b) marginal adjustment, and (c) both regression then marginal adjustment, using the full vector of summary statistics.
Panel (d) shows the same as (c) but using the lower dimensional vector of summary statistics informative for $(B_1,k_1)$. Panel (e) shows the estimate for the Gaussian copula ABC approximation.
\label{gkmodel}  }
\end{figure}

Contour plots of various estimates of the bivariate $(B_1,k_1)$ posterior marginal distribution using ABC rejection sampling are illustrated in Figure \ref{gkmodel}. The top panels
show estimates using the full ($p$-dimensional) vector of summary statistics with (a)  regression adjustment and (b) marginal adjustment, respectively. The performance of each approach individually is poor as the distributions do not exhibit the more accurately estimated dependence structures observed in the remaining panels.      
These estimates are based on ABC rejection sampling with both marginal and regression adjustments, using (c) the full vector of summary statistics, and (d)
the marginally informative summary statistics for $(B_1,k_1)$.
The similarity between panels (c) and (d) indicates that the marginally informative summary statistics are indeed highly informative for the parameter pair  $(B_1,k_1)$. Finally, panel (e) illustrates the Gaussian copula ABC approximation. 
The similarity between panels (d) and (e) indicates that the copula model provides an excellent approximation of the bivariate posterior marginal distribution. 

\subsection{A non-linear multiple quantile regression analysis}
\label{sec:toy examplequantile}

Quantile regression can provide a robust alternative to standard mean regression. \index{Quantile regression} Model estimates obtained at multiple quantile levels can also provide a more complete picture of the conditional distribution between predictor and response. For a regression with a single 
covariate $x$, and response $y$, the linear model corresponding to the $\tau$-th quantile,  $Q_y(\tau|x)$, is given by
\begin{equation*}
Q_{y}(\tau|x)=\alpha_{\tau}+\beta_{\tau}x
\end{equation*}
where the coefficients $\alpha_\tau$ and $\beta_{\tau}$ depend on the quantile level, $\tau\in(0,1)$.
Standard methods fit quantile regressions independently for each quantile level, which can lead to problems of quantiles crossing and a lack of borrowing of information across the quantile levels \cite{rodrigues16}.

Bayesian approaches to quantile regression require the specification of a likelihood. However, exact and tractable likelihood functions are often not available for these models. 
Quantile regression requires the inversion of
many conditional quantile distributions, which are often not analytically available, although numerical grid search can be used (e.g. \shortciteNP{tokdar2012}; \shortciteNP{Reich2010}). However, in the presence of larger data sets, numerical grid searches can become
computationally prohibitive, see for example \shortciteN{Reich2010} who suggests using approximations as an alternative.

\begin{figure}[tb]
\centering
\includegraphics[width=0.7\textwidth]{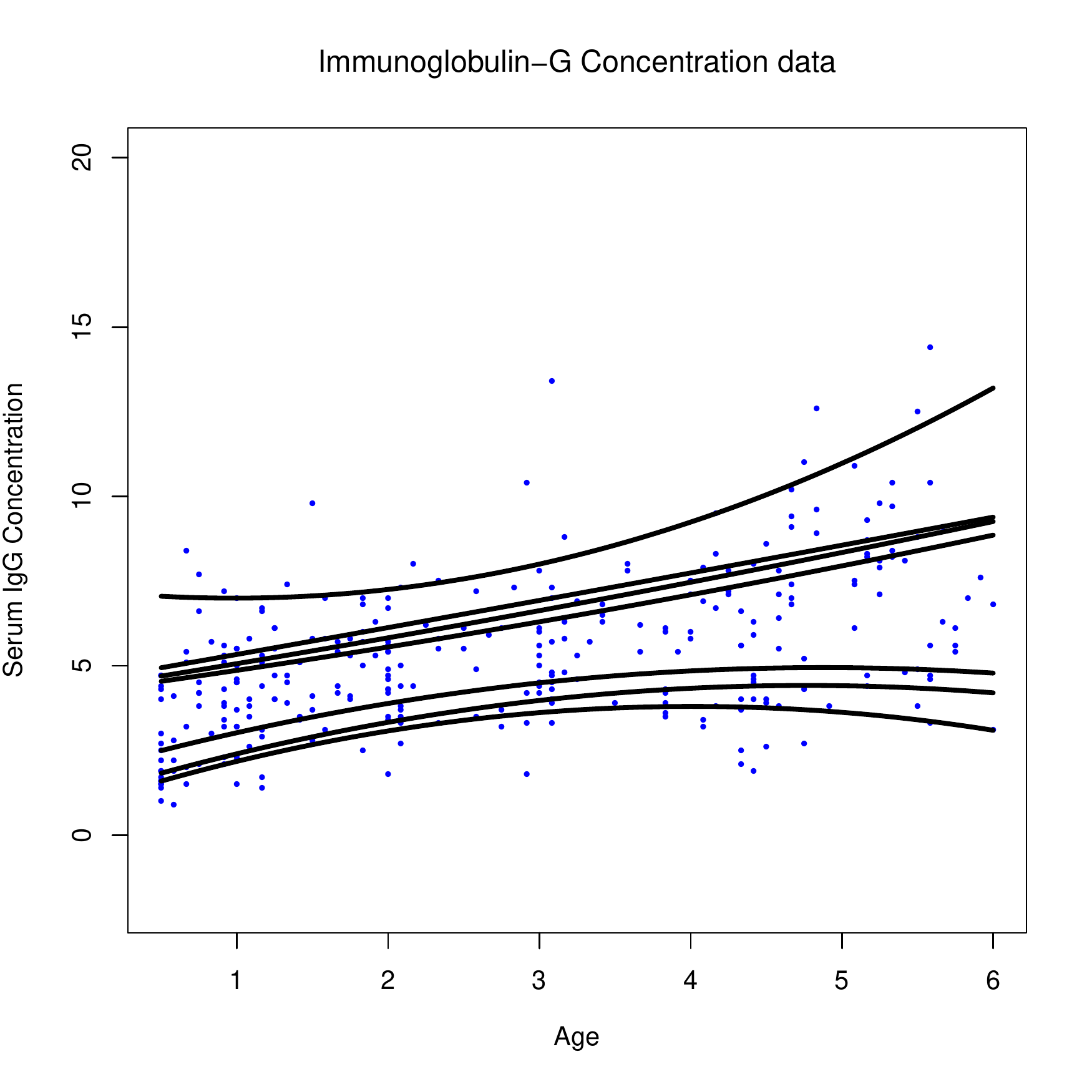}
\caption{\small  The immunoglubulin-G (IgG) dataset. The fitted lines correspond to the classical  quantile estimator at the quantile levels $\tau=0.1, 0.2, 0.3, 0.7, 0.75, 0.8, 0.95$. }
\label{fig:igg}
\end{figure}

We consider a dataset for analysing immunodeficiency in infants. In the search for reference ranges to help diagnose infant immunodeficiency, \shortciteN{Isaac1983} measured the serum concentration of immunoglobulin-G (IgG) in 298 preschool children. We are interested in estimating the IgG conditional quantiles at the  levels $\tau=0.1,0.2,0.3, 0.7, 0.75, 0.8, 0.95$. A quadratic model in age ($x$) is used to fit the data due to the expected smooth change of IgG with age, so that
\begin{equation}
\label{eqn:quantile2}
Q_{y}(\tau|x)=\alpha_{\tau}+\beta_{\tau}x+\eta_{\tau}x^2.
\end{equation}
Figure \ref{fig:igg} illustrates this dataset. The black lines show the separately fitted regression lines for the different quantile levels, based on a frequentist estimator using the {\tt quantreg} package in {\tt R} \cite{Koenker2005}. Since these curves are fitted separately, no correlation is assumed between the quantile curves, and for close quantile levels $\tau$ the fitted quantile estimates can easily cross each other. In practice, strong correlations can exist between curves close to each other, and the true quantile levels will not cross.

We follow the linearly-interpolated likelihood function approach of \shortciteN{fengch2015} as a data model $p(s|\theta)$, while extending their quantile function $Q_{y}(r|x)$ to contain more than one predictor as in (\ref{eqn:quantile2}). For each observed covariate $x_{obs,i}$, $i=1,\ldots,n$, a synthetic data point $y_i$ can be obtained via
$$y_i=Q_{y}(\tau_j|x_{obs,i}) + \frac{Q_{y}(\tau_{j+1}|x_{obs,i}) - Q_{y}(\tau_j|x_{obs,i})  }{\tau_{j+1}-\tau_j}(u_i-\tau_j ),$$
where $u_i\sim U(0,1)$, where $j$ is determined so that
$\tau_j< u_i < \tau_{j+1}$, and where $Q_{\tau}(y|x)$ is the model (\ref{eqn:quantile2}) which depends on parameters $\alpha_{\tau}$, $\beta_{\tau}$ and $\eta_{\tau}$. If $u_i<\tau_1$, $y_i$ is generated from a normal distribution centred on $\bar{y}_{obs}$, with standard deviation 3 times the sample standard deviation of $y_{obs}$, and truncated below $Q_y(\tau_1|x)$. Similarly, if $u_i>\tau_m$, we simulate from the same distribution except that it is truncated above $Q_y(\tau_m|x)$.
The parameters $\alpha_{\tau}, \beta_{\tau}$ and $\eta_{\tau}$ are sampled from multivariate Gaussian prior distributions $\pi(\theta)$, with mean vector and covariance matrix based on the estimates obtained using {\tt quantreg}. 
This prior is constrained to satisfy the quantile monotonicity condition so that the fitted quantile regression lines do not cross.

The full vector of summary statistics is constructed as
 \begin{equation*}
\begin{array}{ll}
s=S(y)=\left( \right.&\hat{\alpha}_{\tau_1},\ldots, \hat{\alpha}_{\tau_m},  \hat{\beta}_{\tau_1}, \ldots, \hat{\beta}_{\tau_m}, \hat{\eta}_{\tau_1}, \ldots, \hat{\eta}_{\tau_m}, \\
& pu_{\tau_1}, \ldots, pu_{\tau_m}, pl_{\tau_1},\ldots, pl_{\tau_m}, q_1(y), \ldots, q_{100}(y) \left.\right)^\top
 \end{array}
 \end{equation*}
where $\hat{\alpha}_{\tau}$, $ \hat{\beta}_{\tau}$ and $ \hat{\eta}_{\tau}$ are the independent frequentist estimators for $\alpha_\tau$, $\beta_\tau$  and $\eta_\tau$ at  quantile level $\tau$, $pu_{\tau}$ is the proportion of data points above the $\tau$th quantile curve, $pl_{\tau}$ is the proportion of data points below 
the $\tau$th quantile curve, and $q_1(y), \ldots, q_{100}(y)$ are the 100 equally
spaced quantiles of the data $y$.  The summary statistics for $\alpha_{\tau_1}$ are $\hat{\alpha}_{\tau_i}, pu_{\tau_i}, pl_{\tau_i}$ and the closest 20 quantiles $q_1(y), \ldots, q_{100}(y)$ to the level $\tau_i$. Similarly, for $\beta_{\tau_j}$, the marginally
informative summary statistics will be $\hat{\beta}_{\tau_j}, pu_{\tau_j}, pl_{\tau_j}$ and the closest 20 quantiles $q_1(y), \ldots, q_{100}(y)$ to the level $\tau_j$; and so on. Then for the summaries of the bivariate margin, $(\alpha_{\tau_i}, \beta_{\tau_j})$, we concatenate the two sets of summaries.

The following analysis is based on $N=1,000,000$ samples $(s^{(\ell)},\theta^{(\ell)})\sim p(s|\theta)\pi(\theta)$, $\ell=1,\ldots,N$.
We specify the smoothing kernel $K_h(\cdot)$ as uniform over the range $(-h,h)$ and determine $h$ as the 0.001 quantile of the Euclidean distances between observed and simulated summary statistics. Our model simultaneously fits the seven quantile levels shown in Figure \ref{fig:igg}, resulting in a $p=21$ dimensional model with $q=135$ total summary statistics. Note that with  the application of post-hoc adjustments,
monotonicity of the conditional quantiles may not be preserved. 
If this occurs, the offending samples may simply be discarded, although a preferable solution is the development of adjustments that flexibly respect constraints.

\begin{figure}[tbh]
\centering
\includegraphics[width=5.8cm]{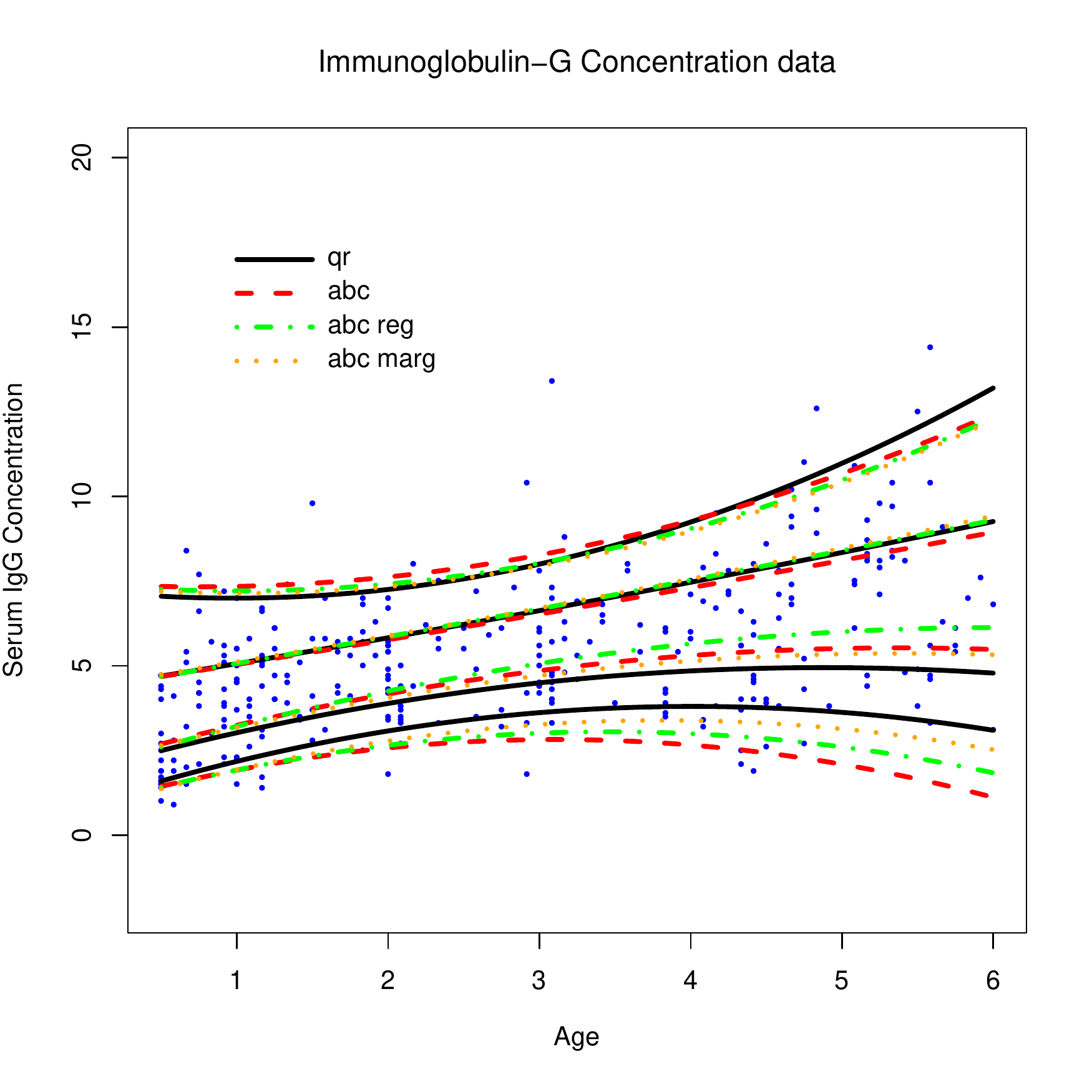}
\includegraphics[width=5.8cm]{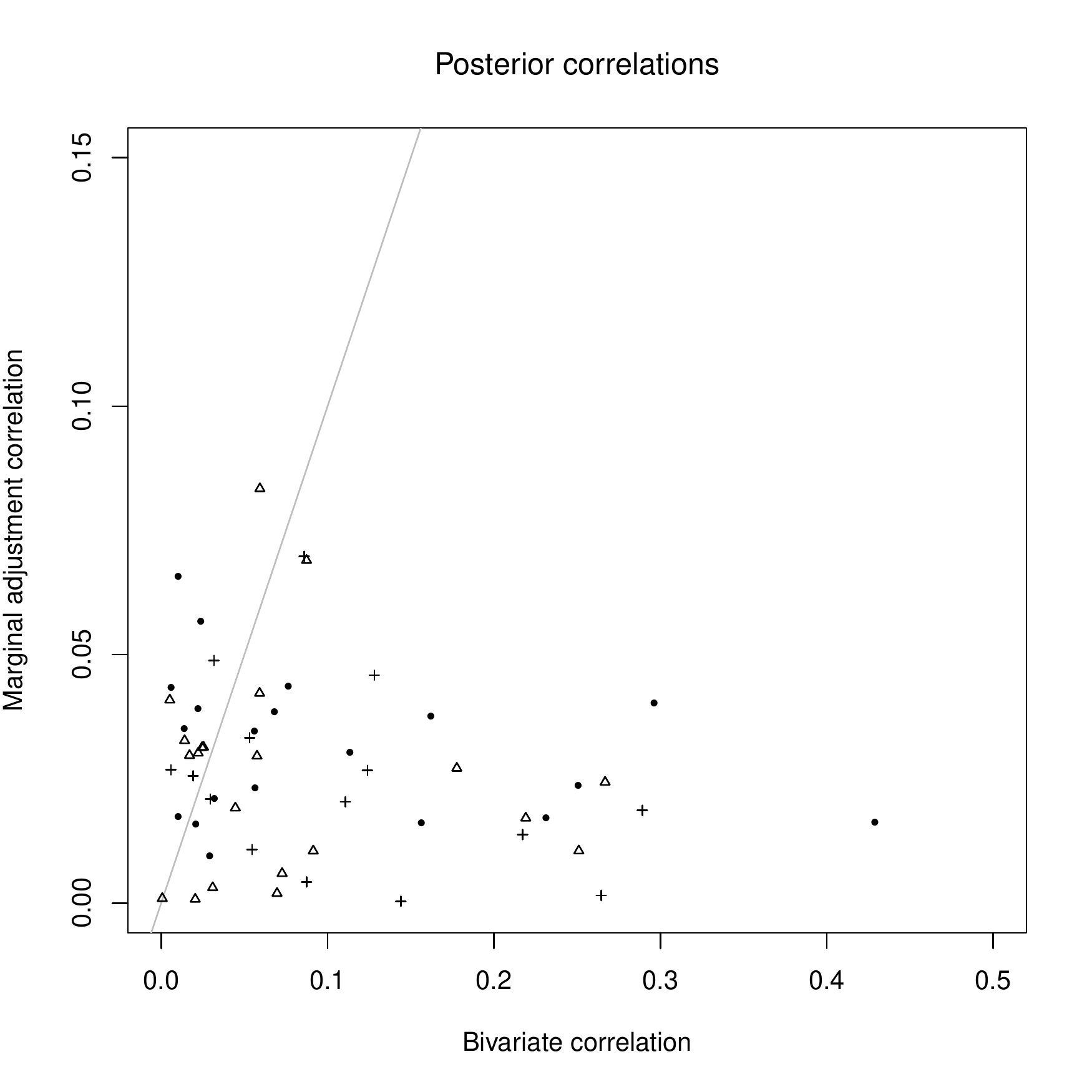}
\caption{\small Left panel: Posterior mean predictive conditional quantile estimates using the full vector of summary statistics based on standard ABC (dashed line), regression adjusted ABC (dot-dash line), and both regression and marginally adjusted ABC, for the quantiles $\tau=0.1, 0.2, 0.3, 0.7, 0.75, 0.8, 0.95$. For clarity only $\tau=0.1,0.3,0.75,0.95$ level quantiles are shown. 
Right panel: Estimated correlation of posterior margins $\pi(\alpha_i,\alpha_j|s_{obs})$ (dot), $\pi(\beta_i,\beta_j|s_{obs})$ (triangle) and $\pi(\eta_i,\eta_j|s_{obs})$ (plus) $i\neq j,$ for regression and marginally adjusted ABC with the full vector of summary statistics ($y$-axis), against that for Gaussian copula ABC ($x$-axis).}
\label{fig:iggcor}
\end{figure}

Figure \ref{fig:igg} (left panel) shows the mean predicted conditional quantile estimates for the levels $\tau=0.1, 0.3, 0.75, 0.95$ based on fitting the seven quantile level model. 
Although the true quantile curves are not known here, we might expect the independently fitted frequentist estimates to provide a reasonable guide to the truth in this analysis. When the sample size is reasonably large (here $n=298$), the frequentist approach can produce estimators with good properties (such as a reduced chance for neighbouring quantiles to overlap as $n$ gets large).  As a result, in the current example, the frequentist estimates should be expected to produce similar results to the Bayesian approaches, particularly in the non-extreme regions where there is more data. However the Bayesian analyses naturally enforce non-crossing of quantiles, and so are preferable for this reason, in spite of the approximate posterior. Results from three different ABC variants are shown in Figure \ref{fig:iggcor} (left panel).
For most quantile levels there are small differences between the marginal univariate quantile estimates, although quantile non-crossing is enforced in each of the Bayesian estimates. For the lower $\tau=0.1$ quantile where data is more scarce, increasing the quality of the ABC posterior approximation from standard rejection ABC (dashed line) to regression adjusted ABC (dot-dash line) to regression and marginally adjusted ABC (dotted line), produces a marginal quantile that is increasingly close to the frequentist estimate, and which roughly partitions 10\% of the data below it. This suggests that there is some ABC approximation error (although this is less obvious in the upper $\tau=0.95$ quantile), but that this is less apparent the better the ABC approximation becomes.

In the case of these marginal quantile estimates, Gaussian copula ABC produces quantile estimates (not shown) that are highly similar to the regression and marginally adjusted estimates (dotted line). However, the real differences here are in the quality of the dependence structure of the ABC posterior.
 Figure \ref{fig:iggcor} (right panel)  shows the correlation in the estimated posterior bivariate margins of $(\alpha_i,\alpha_j)$, $(\beta_i,\beta_j)$ and $(\eta_i,\eta_j)$ for $i\neq j$ when using Gaussian copula ABC ($x$-axis) and standard ABC with regression and marginal adjustment using the full vector of summary statistics ($y$-axis). Here
it is evident that Gaussian copula ABC is able to capture correlations in the bivariate margins that are missed by regular ABC, even when using the univariate marginal adjustment. 
The quality of the posterior approximation will be vital when considering analyses that critically depend on  full, multiple quantile inference. This lends support to
 the Gaussian copula approach as a viable ABC model approximation able to capture much of the bivariate dependence structure of $\pi(\theta|s_{obs})$.

\section{ABC approximation of the sampling distribution of summary statistics}
\label{sec:approxLik}

An alternative to direct ABC approximation of the posterior distribution $\pi(\theta|s_{obs})$ is to instead approximate the sampling distribution of summary statistics $p(s|\theta)$
(\shortciteNP{leuenberger+w10,fan+ns13}), thereby approaching the intractable likelihood problem from the more usual ABC conditional density estimation perspective.
The resulting estimated density is then an analytically tractable approximation of the likelihood function for a Bayesian 
analysis using conventional Bayesian computational tools.  Such approaches may be preferable in problems where inference is required for 
multiple datasets arising from the same model.

One way to achieve this is to first
estimate the joint distribution of $(s,\theta)$ flexibly and to then condition on observing $s=s_{obs}$ in the joint model.  This approach was considered by \shortciteN{bonassi+yw11} using multivariate normal mixture 
models for the density estimator on $(s,\theta)$.  
Synthetic  likelihood \cite{wood10} is another method that directly approximates  the likelihood via an assumed density such as $p(s|\theta)\approx N_q(\mu(\theta),\Sigma(\theta))$ where the mean $\mu(\theta)$ and covariance matrix $\Sigma(\theta)$ are unknown functions of the parameter $\theta$. Various techniques are then needed to estimate $\theta$. For further details on synthetic likelihoods see e.g. \citeN{wood10}, \shortciteN{wood18} and \shortciteN{drovandi+gkr18}.

\subsection{A flexible regression density estimator}

We describe the flexible conditional density estimation approach of \shortciteN{fan+ns13}.  \index{Regression density estimation}
As with other ABC density estimators, it is constructed from a sample of $N$ summary statistic and parameter pairs
$(s^1,\theta^1),\dots,(s^N,\theta^N)$ drawn from a distribution $p(s|\theta)h(\theta)$.  Note that while the summary statistics are generated given $\theta$ from the sampling distribution for the intractable model of interest, the parameters are not necessarily generated from the prior.  Instead, $h(\theta)$ is a distribution
chosen to reflect the region over which the likelihood should be well approximated.  Some rough knowledge of the high likelihood region of the parameter space, 
perhaps based on an initial pilot analysis, is useful for setting $h(\theta)$.  The method of
\shortciteN{fan+ns13} is based on relating
the summary statistics $s$ 
to $\theta$ 
by regression approximations, and so it is useful if the actual relationships between $s$ and $\theta$ are as simple as possible. One convenient procedure to achieve this is the
semi-automatic summary statistic
approach of \citeN{fearnhead+p12} which constructs one summary statistic per parameter where each summary statistic is an estimate of the posterior mean value of the parameter, based on a pilot run.  
That is, $s_k$ is the univariate summary statistic informative for $\theta_k$, $k=1,\ldots, p$, with 
$s^j=(s^j_1,\dots,s^j_p)$.  

The first step 
is to build marginal regression models for each component of $s$ conditional on $\theta$.  The training data $(s_k^1,\theta^1),\dots,(s_k^N,\theta^N)$ is used to build the marginal model for
$s_k$ resulting in an estimated marginal density $\hat{f}_k(s_k|\theta)$ for $s_k$.  \shortciteN{fan+ns13} use a fast variational method for fitting mixture of heteroscedastic regression models (\shortciteNP{nott+tvk12,tran+nk12})  
for the conditional density estimation.  

Then a conditional density estimate for the joint distribution of $s$ given $\theta$ is constructed, using a method closely related to that 
considered in \shortciteN{giordani+mtk13} for the unconditional case.  The data $(s^j,\theta^j)$ are transformed to $(U^j,\theta^j)$, where $U_k^j=\Phi^{-1}(\hat{F}_k(s_k^j|\theta^j))$, 
where $\hat{F}_k(s_k|\theta)$ is the distribution function corresponding to the density $\hat{f}_k(s_k|\theta)$.  If the marginal densities for each
$s_k$ are well estimated, the transformation to $U^j$ makes each component of $U^j$ approximately standard normal regardless of the value of $\theta$.  
A mixture of normals model is then fitted to the data $(U^j,\theta^j)$, $j=1,\dots,N$.  Write the fitted normal mixture as 
$$\sum_{k=1}^K w_k N(\mu_k,\Psi_k)$$
where $N(\mu,\Psi)$ denotes the multivariate normal distribution with mean $\mu$ and covariance matrix $\Psi$, $(\mu_k,\Psi_k)$, $k=1,\dots,K$ are means
and covariances of $K$ normal mixture components, and $w_k$, $k=1,\dots,K$ are mixing weights, $w_k\geq 0$, $\sum_{j=1}^Kw_j=1$.  The mixture model
for the joint distribution of $(U,\theta)$ then implies a normal mixture model for the conditional density of $U|\theta$, 
$$\sum_{k=1}^K w_k^c N(\mu_k^c,\Psi_k^c)$$
where
$$w_k^c=\frac{w_k \phi(\theta;\mu_k,\Psi_k)}{\sum_{j=1}^K w_j \phi(\theta; \mu_j,\Psi_j)}$$
are mixing weights with $\phi(\theta;\mu,\Psi)$ denoting the multivariate normal density function in $\theta$ with mean $\mu$ and covariance matrix $\Psi$, 
and $\mu_k^c$ and $\Psi_k^c$ are the conditional mean and covariance of $U$ given $\theta$ in the $k$-th multivariate normal component
$N(\mu_k,\Psi_k)$ in the joint mixture model.  Write $\hat{g}(U|\theta)$ for the resulting estimated conditional density of $U$ given $\theta$.  
Inverting the transformation of $s$ to $U$ then produces an estimate of the conditional density of $s$ given $\theta$, 
\begin{align}
 \hat{L}(s|\theta) & =\hat{g}(U|\theta)\prod_{j=1}^K \frac{\hat{f}_j(s_j|\theta)}{\phi(U_j;0,1)}.  \label{lapprox}
\end{align}
An approximation of the observed data likelihood is then given by $\hat{L}(s_{obs}|\theta)$.

The purpose of the transformation from $s$ to $U$ is to simplify the mixture modelling of the joint distribution $(U,\theta)$ compared to what would be required
to estimate the joint distribution of $(s,\theta)$.  Note that in $\hat{L}(s|\theta)$ the marginal density of $s_k$ is not exactly $\hat{f}(s_k|\theta)$ due
to the fact that the estimated marginal distributions in $\hat{g}(U|\theta)$ are not exactly standard normal.  
\shortciteN{giordani+mtk13} suggest replacing the $\phi(U^j;0,1)$  in (\ref{lapprox}) by its exact marginal distribution in $\hat{g}(U|\theta)$, but \shortciteN{fan+ns13} found that
good approximations to $L(s_{obs}|\theta)$ were obtained without this step.  

The above conditional density estimation method seeks to estimate each univariate marginal conditional distribution $s_k|\theta$ arbitrarily well, while approximating the overall joint dependence structure by a mixture of normals model. This approach can work well in relatively high dimensions, in the order of tens to hundreds, provided that the dependence structure is relatively straightforward to capture. This also underlines the importance of techniques that can produce summary statistics with simple relationships to $\theta$, such as the method developed by \citeN{fearnhead+p12}.

\subsection{Analysis of stereological extremes} \index{Stereological extremes}

To illustrate the regression density estimation approach we reanalyse a dataset originally analysed using ABC methods by \shortciteN{bortot+cs07}, and which was previously considered in \shortciteN{sisson+fb18}.  The data comprise information about the intensity and size distribution of inclusions
 in a 3 dimensional block of clean steel, with the recorded observations being the inclusion sizes (above a threshold of $\nu_0=5\mu$m), and their number, observed in a 2-dimensional cross-section.

\shortciteN{bortot+cs07} considered models assuming spherical or ellipsoidal inclusion shapes. For the elliptical model the inclusion size is the length of the major axis of the two-dimensional planar ellipse.
In both models the locations of the inclusions above 5$\mu$m in size follow a Poisson process with intensity $\lambda$.  
Conditional on having an inclusion larger than $\nu_0$, the distribution of the inclusion size is generalized Pareto, with scale parameter $\sigma>0$
and shape parameter $\xi$.  So in both models there
are 3 parameters, $\theta=(\lambda,\sigma,\xi)^\top$.  For the analysis the priors are $\log \lambda\sim N(0,100^2)$, $\sigma\sim \mbox{Gamma}(0.01,0.0001)$ and
$\xi\sim N(0,100^2)$.  For the spherical inclusion model it is possible to directly evaluate the likelihood, but for the ellipsoidal inclusion model this is not possible and so
ABC methods are an attractive option.  Here we focus on the ellipsoidal inclusion model.
An analysis of standard rejection ABC with regression adjustment for the spherical model can be found in \citeN{erhardt+s16}.

The high-dimensionality aspect of this analysis comes from the number of summary statistics, rather than the number of parameters.
The summary statistics used comprise the logarithm of the number of inclusions observed in the two-dimensional cross-section ($s_1=111$), and $s_{j+1}=\log (q_{(j+1)}-q_{(j)})$, $j=1,\dots,111$, 
where the $q_{(j)}$, $j=1,\dots,112$ are 112 equally spaced quantiles of the observed inclusion sizes.  This gives $q=112$ summary statistics in total, and corresponds to conditional density estimation in 112+3=115 dimensions.

The conditional density estimation method requires the choice of $h(\theta)$. This is achieved via a pilot analysis by firstly sampling values $(s^{i},\theta^{i})$, $i=1,\dots,n$, 
 where the $\theta^{i}$ are sampled from a uniform distribution over a range wide enough to include the support of the posterior and
the $s^{i}$ are sampled from $p(s|\theta^i)$.  
The sample mean $\hat{\mu}$ and covariance matrix  $\hat{\Sigma}$ are then calculated for those $\theta$ values for which 
$\|s^{i}-s_{obs}\|\leq 20$.
The distribution $h(\theta)$ is then specified as the truncated normal distribution
$$h(\theta)\propto N(\hat{\mu},\hat{\Sigma})I((\theta-\hat{\mu})^\top\hat{\Sigma}^{-1}(\theta-\hat{\mu})<9).$$
The conditional density estimation method for estimating $p(s|\theta)$ is then implemented using $N=5000$ draws $(s^i,\theta^i)\sim p(s|\theta)h(\theta)$, $i=1,\ldots,N$.

\begin{figure}
\includegraphics[scale=0.55]{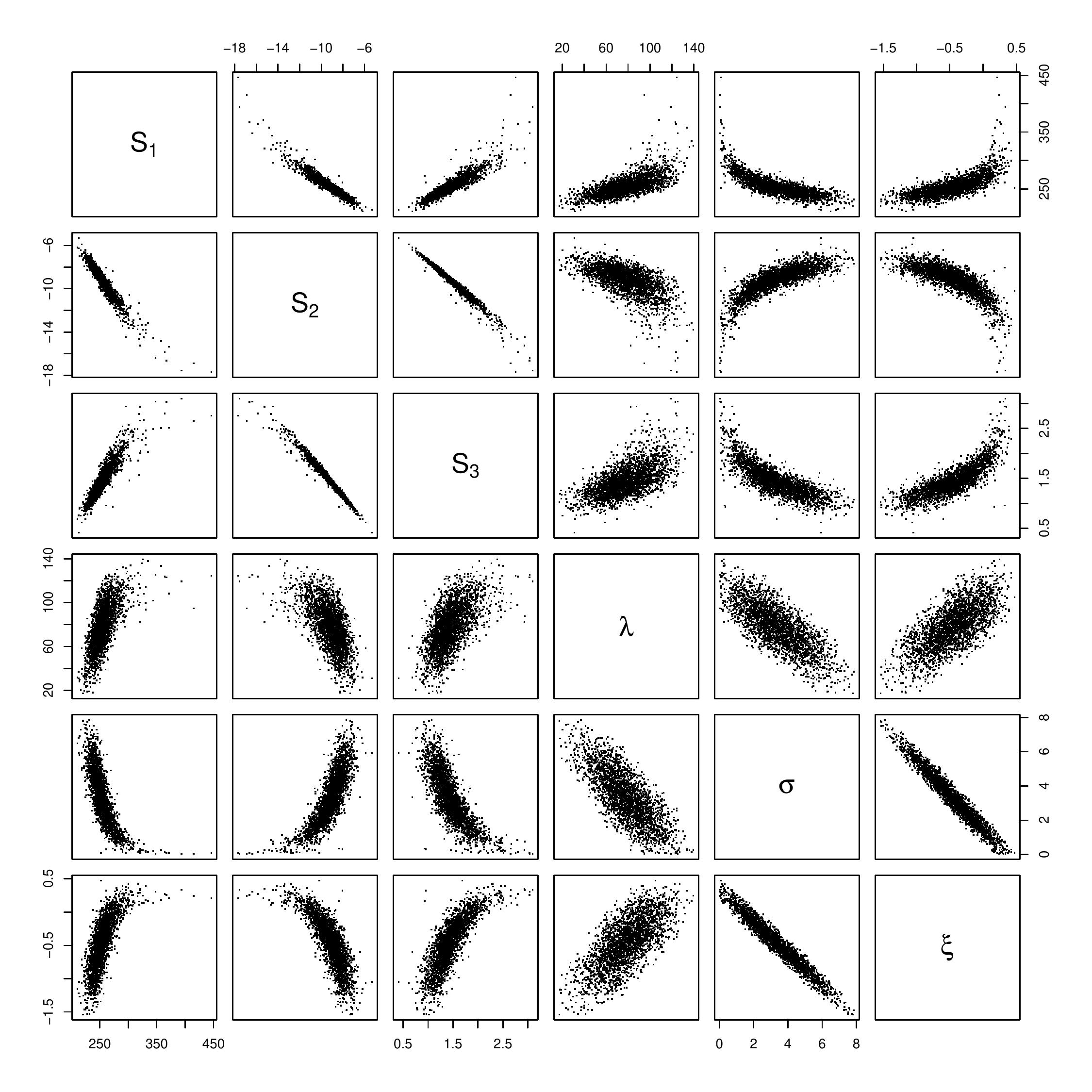}
\caption{\small Pairwise scatterplots between the Fearnhead and Prangle (2012) semi-automatic summary statistics $s_1$, $s_2$ and $s_3$ and the parameters $\lambda$, $\sigma$ and $\xi$ for the ellipsoidal inclusions model. \label{Stheta} }
\end{figure}

\begin{figure}
\includegraphics[scale=0.55]{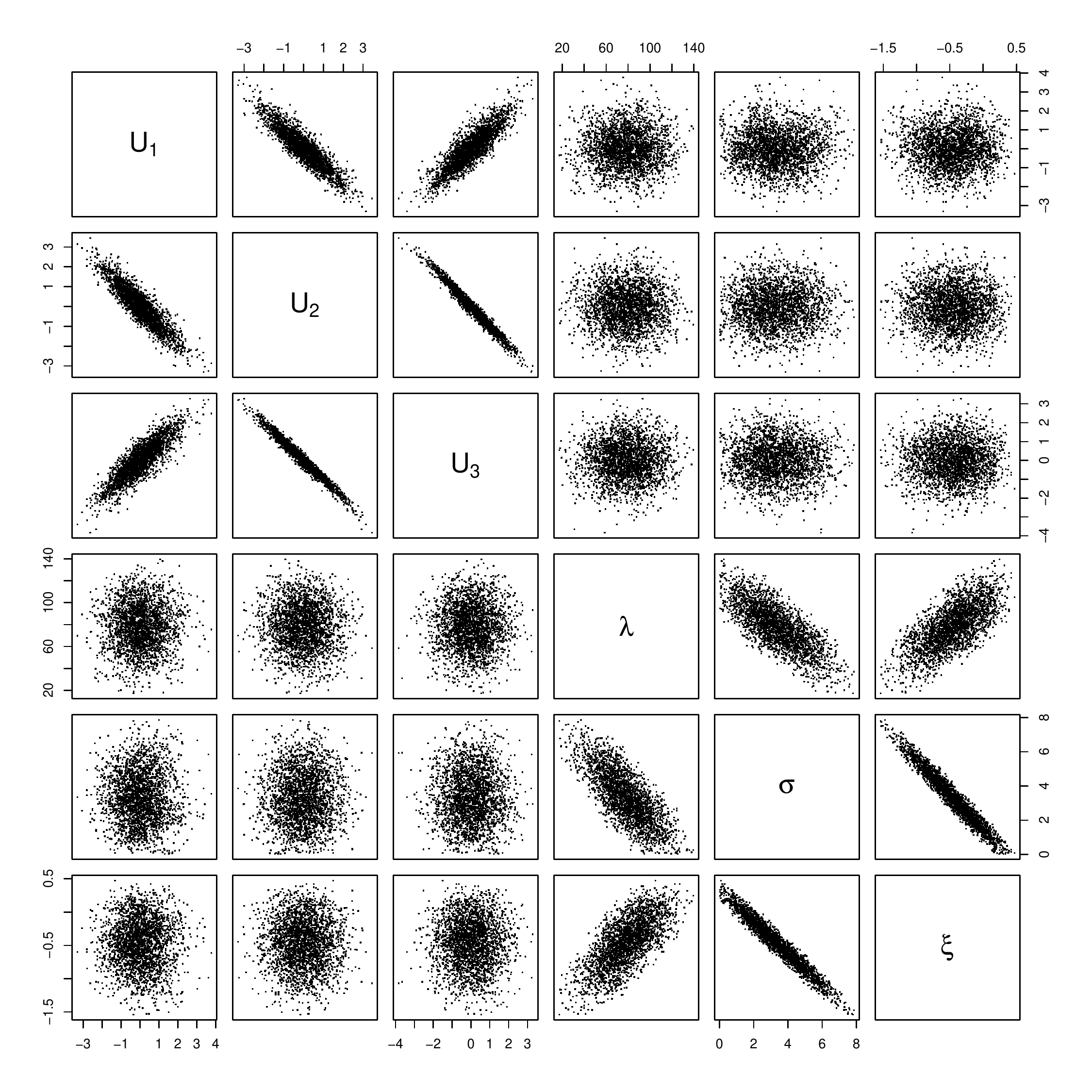}
\caption{\small Pairwise scatterplots between the transformed Fearnhead and Prangle (2012) summary statistics $U_1$, $U_2$ and $U_3$ and the parameters  $\lambda$, $\sigma$ and $\xi$ fo the ellipsoidal inclusions model. \label{Utheta} }
\end{figure}

For comparison with the 115-dimensional regression density estimation approach, an additional analysis is performed in only 6-dimensions, using the 3-dimensional summary statistics obtained using
the semi-automatic method of 
\citeN{fearnhead+p12}.   
Figure \ref{Stheta} shows pairwise scatterplots of the components of $s$ and $\theta$ for the samples generated from $h(\theta)p(s|\theta)$ (plotting the \citeNP{fearnhead+p12} statistics analysis for clarity).  
The resulting scatterplots
after fitting the flexible models $f_k(s|\theta)$ to 
the univariate marginal distributions and transforming to the statistics $U$ are illustrated in Figure \ref{Utheta}.  
Clearly the dependence structure
has been greatly simplified, which facilitates the accurate mixture modelling of $(U,\theta)$.

\begin{figure}
\centering
\includegraphics[scale=0.40]{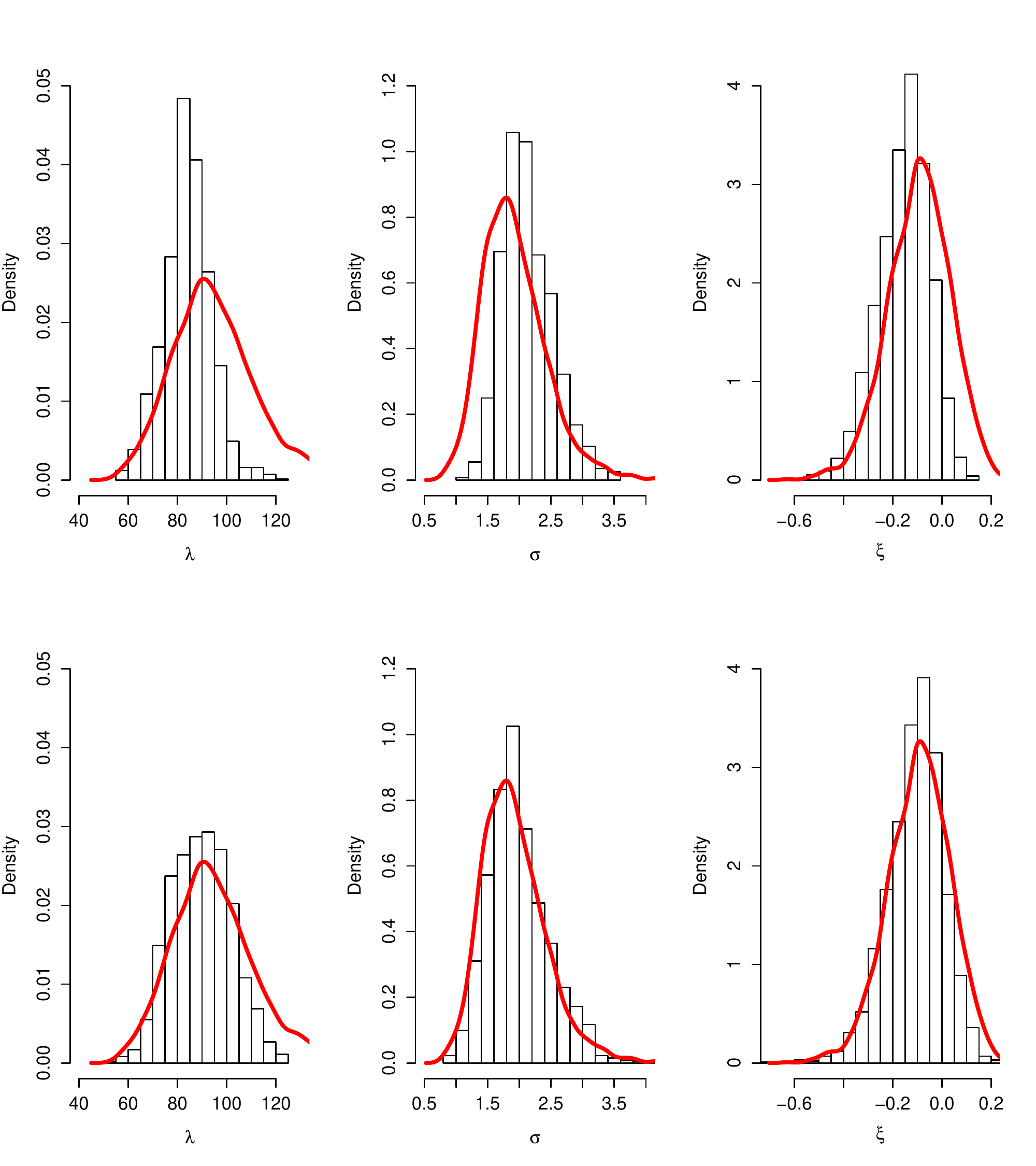}
\caption{\small Histograms illustrating the estimated marginal posterior distributions obtained by regression density estimation for the ellipsoidal inclusions model using 112 summary statistics (top rows) and the 3 Fearnhead and Prangle (2012) summary statistics (bottom rows). The solid line shows the `gold standard' marginal densities obtained using the method of Bortot et al. (2007),  
with a kernel scale parameter of $h=0.33$}. \label{Elliptical_histogram} 
\end{figure}

The histograms in Figure \ref{Elliptical_histogram} show the regression density estimated
marginal posterior distributions obtained by using
the original 112 summary statistics (top panels), and the lower dimensional \citeN{fearnhead+p12} statistics (bottom panels). The solid line illustrates the density estimates obtained by the `gold standard' ABC-MCMC analysis
of \shortciteN{bortot+cs07}  using large computational overheads.
It is apparent that even when modelling the original high-dimensional set of summary statistics, reasonable answers are obtained using the regression density approach, although using the same method but with
the \citeN{fearnhead+p12} summary statistics naturally results in an improved performance.  

\section{Other approaches to high-dimensional ABC}

Beyond the density estimation techniques described above, there are a few alternative approaches for extending ABC analyses to higher dimensions.
ABC methods have been previously developed for  functional parameters, specifically in the case of non-parametric hierarchical density estimation \shortcite{rodrigues+ns16}. However, while these `infinite-dimensional' parameters require the development of specialised ABC methods (such as a functional regression adjustment), the dimensionality of these techniques is strictly not high-dimensional in the sense considered in this chapter.

Various possibilities are available when the model of interest has a known and exploitable structure.
The simplest of these is where the model factorises into a hierarchical structure $p(s|\theta,\phi)=f(\theta|\phi)\prod_i p_i(s^{(i)}|\theta^{(i)})$ (e.g. \shortciteNP{bazin+db10}), where $s^{(i)}$ and $\theta^{(i)}$ denote mutually exclusive partitions of $s$ and $\theta$. In this case, the ABC approximation to the joint posterior $\pi(\theta|s_{obs})$ may be naturally constructed using the lower dimensional comparisons $\|s^{(i)}-s^{(i)}_{obs}\|$ only.

When the data model can be written in a conditional factorisation form\\ $p(s|\theta) = p(s_1|\theta)\prod_{i=1}^q p(s_i|s_{1:(i-1)},\theta)$, where $s_{1:k}=(s_1,\ldots,s_k)^\top$, and where conditional simulation from $p(s_i|s_{1:(i-1)},\theta)$ is possible, \shortciteN{barthelme+c14} (see also \shortciteNP{white+kp15}) proposed an expectation-propogation ABC scheme.
If $s_i$ is low dimensional then $p(\theta|s_{i,obs},s_{1:(i-1),obs})$ (that is, the posterior obtained by matching $\|s_i-s_{i,obs}\|$ based on simulating conditionally on $s_{1:(i-1),obs}$) can be well estimated via regular ABC, which \shortciteN{barthelme+c14} then approximate by a Gaussian density. \index{Expectation propagation} This leads to a Gaussian approximation of $p(\theta|s_{obs})$, which may be accurate if the number of summary statistics is large. This may be realistic if
the summary statistics $S(y)=y$ are the observed data.
See \shortciteN{barthelme+cc17} (this volume) for further details of this approach.

\shortciteN{kousthanas+lhqfw15} consider constructing an MCMC-sampler with univariate updates to sample from each univariate conditional distribution $\pi(\theta_i|\theta_{-i},s_{obs})$. Here they note that if a low-dimensional summary statistic can be identified that is sufficient for the conditional distribution of $\theta_i|\theta_{-i}$, then an ABC-MCMC sampler can be implemented that compares summary statistics of a much lower dimension than than the full vector $s$ at each update step. They demonstrate this approach on a high-dimensional linear model with univariate summary statistics for each parameter update.

Finally, synthetic likelihood methods were discussed in Section \ref{sec:approxLik} as a method to approximate the likelihood function using an assumed parametric form e.g. $p(s|\theta)\approx N_q(\mu(\theta),\Sigma(\theta))$ \cite{wood10}. \index{Synthetic likelihood} As this technique relies on estimating $\mu(\theta)$ and $\Sigma(\theta)$ for each $\theta$ based on a potentially large number of Monte Carlo samples from $p(s|\theta)$, this approach can have high computational overheads. Variational Bayes has only recently been considered as a possible approach for fitting intractable models with synthetic likelihoods, but with greatly reduced computational costs. This then allows higher dimensional analyses to be implemented. See e.g. \shortciteN{tran+nk16} and \shortciteN{ong+ntsd16} for further details on this technique.

\section{Discussion}\label{disc}

Given that a direct ABC approximation of the joint posterior distribution $\pi(\theta|s_{obs})$ involves a kernel density approximation of the likelihood, where the dimensionality involved is the dimension of the summary statistic $s$, it might initally seem that development of useful, general purpose methods for high-dimensional ABC may not be possible.  However, if we are prepared to step away from the limiting comparison of $\|s-s_{obs}\|$ within the likelihood approximation of standard ABC methods, and build an approximations to $\pi(\theta|s)$  or $p(s|\theta)$ from approximations of lower dimensional distributions,
then it may be possible to develop useful ABC posterior approximations even in high dimensional settings.  The key idea in these approaches is that instead of matching a single vector of summary statistics in high dimensions, $\|s-s_{obs}\|$, we instead match many different low dimensional summary statistic vectors in constructing our joint posterior approximation. 
While the methods described here will not always work for posterior distributions with a highly complex dependence structure, or in very high dimensions,
further development 
of related methods using the same ``divide and conquer" strategy may be a promising direction for future research in high-dimensional ABC.    
This may be particularly true for those methods that are easily parallelisable in their implementation.

Another area that perhaps has good potential for future research involves those techniques related to pseudo-marginal MCMC methods (see \shortciteNP{andrieu+lv17}), which is currently seeing a surge of research interest beyond ABC. These methods have opened up ways to perform exact estimation and sampling for models with intractable likelihood functions, the ideas of which can be extended to implement various forms of approximation of posterior distributions.  These include synthetic likelihoods (see \shortciteNP{drovandi+gkr18}) and variational Bayes methods, which can both be fast to implement, and for which the latter tends to underestimate uncertainty.
The extension of likelihood-free inference methods to problems of higher dimension is a very active research area and promises to be so for the forseeable future.

\section*{Acknowledgements}

SAS is supported by the Australian Research Council under the Discovery Project scheme (DP160102544), and the Australian Centre of Excellence in Mathematical and Statistical Frontiers (CE140100049).

\bibliographystyle{chicago}
\bibliography{bibtex_example}

\begin{thebibliography}{}

\bibitem[\protect\citeauthoryear{Allingham, King, and Mengersen}{Allingham
  et~al.}{2009}]{allinghamkm09}
Allingham, D.~R., A.~R. King, and K.~L. Mengersen (2009).
\newblock Bayesian estimation of quantile distributions.
\newblock {\em Statistics and Computing\/}~{\em 19}, 189--201.

\bibitem[\protect\citeauthoryear{Andrieu, Lee, and Vihola}{Andrieu
  et~al.}{2018}]{andrieu+lv17}
Andrieu, C., A.~Lee, and M.~Vihola (2018).
\newblock Theoretical and methodological aspects of {MCMC} computations with
  noisy likelihoods.
\newblock In S.~A. Sisson, Y.~Fan, and M.~A. Beaumont (Eds.), {\em Handbook of
  Approximate Bayesian Computation}.

\bibitem[\protect\citeauthoryear{Barber, Voss, and Webster}{Barber
  et~al.}{2015}]{barber+vw15}
Barber, S., J.~Voss, and M.~Webster (2015).
\newblock The rate of convergence for approximate {B}ayesian computation.
\newblock {\em Electronic Journal of Statistics\/}~{\em 9}, 80--105.

\bibitem[\protect\citeauthoryear{Barthelme and Chopin}{Barthelme and
  Chopin}{2014}]{barthelme+c14}
Barthelme, S. and N.~Chopin (2014).
\newblock Expectation propagation for likelihood-free inference.
\newblock {\em Journal of the American Statistical Association\/}~{\em 109},
  315--333.

\bibitem[\protect\citeauthoryear{Barthelm\'e, Chopin, and Cottet}{Barthelm\'e
  et~al.}{2018}]{barthelme+cc17}
Barthelm\'e, S., N.~Chopin, and V.~Cottet (2018).
\newblock Divide and conquer in {ABC}: {Expectation-Propagation} algorithms for
  likelihood-free inference.
\newblock In S.~A. Sisson, Y.~Fan, and M.~A. Beaumont (Eds.), {\em Handbook of
  Approximate Bayesian Computation}. Chapman and Hall/CRC Press.

\bibitem[\protect\citeauthoryear{Bazin, Dawson, and Beaumont}{Bazin
  et~al.}{2010}]{bazin+db10}
Bazin, E., K.~Dawson, and M.~A. Beaumont (2010).
\newblock Likelihood-free inference of population structure and local
  adaptation in a {Bayesian} hierarchical model.
\newblock {\em Genetics\/}~{\em 185}, 587--602.

\bibitem[\protect\citeauthoryear{Beaumont, Zhang, and Balding}{Beaumont
  et~al.}{2002}]{beaumont+zb02}
Beaumont, M.~A., W.~Zhang, and D.~J. Balding (2002).
\newblock Approximate {Bayesian} computation in population genetics.
\newblock {\em Genetics\/}~{\em 162}, 2025--2035.

\bibitem[\protect\citeauthoryear{Biau, C\'{e}rou, and Guyader}{Biau
  et~al.}{2015}]{biau+cg15}
Biau, G., F.~C\'{e}rou, and A.~Guyader (2015).
\newblock New insights into approximate {B}ayesian computation.
\newblock {\em Ann. Inst. H. Poincar\'{e} Probab. Statist.\/}~{\em 51\/}(1),
  376--403.

\bibitem[\protect\citeauthoryear{Blum}{Blum}{2010}]{blum10}
Blum, M. G.~B. (2010).
\newblock Approximate {Bayesian} computation: a non-parametric perspective.
\newblock {\em Journal of the American Statistical Association\/}~{\em 105},
  1178 -- 1187.

\bibitem[\protect\citeauthoryear{Blum and Fran\c{c}ois}{Blum and
  Fran\c{c}ois}{2010}]{blum+f10}
Blum, M. G.~B. and O.~Fran\c{c}ois (2010).
\newblock Non-linear regression models for approximate {Bayesian} computation.
\newblock {\em Statistics and Computing\/}~{\em 20}, 63--75.

\bibitem[\protect\citeauthoryear{Blum, Nunes, Prangle, and Sisson}{Blum
  et~al.}{2013}]{blum+nps13}
Blum, M. G.~B., M.~A. Nunes, D.~Prangle, and S.~A. Sisson (2013).
\newblock A comparative review of dimension reduction methods in approximate
  {Bayesian} computation.
\newblock {\em Statistical Science\/}~{\em 28}, 189--208.

\bibitem[\protect\citeauthoryear{Bonassi, You, and West}{Bonassi
  et~al.}{2011}]{bonassi+yw11}
Bonassi, F.~V., L.~You, and M.~West (2011).
\newblock Bayesian learning from marginal data in bionetwork models.
\newblock {\em Statistical Applications in Genetics and Molecular
  Biology\/}~{\em 10\/}(1).

\bibitem[\protect\citeauthoryear{Bortot, Coles, and Sisson}{Bortot
  et~al.}{2007}]{bortot+cs07}
Bortot, P., S.~G. Coles, and S.~A. Sisson (2007).
\newblock Inference for stereological extremes.
\newblock {\em Journal of the American Statistical Association\/}~{\em 102},
  84--92.

\bibitem[\protect\citeauthoryear{Drovandi, Grazian, Mengersen, and
  Robert}{Drovandi et~al.}{2018}]{drovandi+gkr18}
Drovandi, C.~C., C.~Grazian, K.~Mengersen, and C.~P. Robert (2018).
\newblock Approximating the likelihood in approximate {Bayesian} computation.
\newblock In S.~A. Sisson, Y.~Fan, and M.~A. Beaumont (Eds.), {\em Handbook of
  Approximate Bayesian Computation}. Chapman {\&} Hall/CRC Press.

\bibitem[\protect\citeauthoryear{Drovandi and Pettitt}{Drovandi and
  Pettitt}{2011}]{drovandi+p11}
Drovandi, C.~C. and A.~N. Pettitt (2011).
\newblock Likelihood-free {Bayesian} estimation of multivariate quantile
  distributions.
\newblock {\em Computational Statistics and Data Analysis\/}~{\em 55},
  2541--2556.

\bibitem[\protect\citeauthoryear{Erhardt and Sisson}{Erhardt and
  Sisson}{2016}]{erhardt+s16}
Erhardt, R. and S.~A. Sisson (2016).
\newblock Modelling extremes using approximate {Bayesian} computation.
\newblock In D.~Dey and J.~Yan (Eds.), {\em Extreme Value Modelling and Risk
  Analysis}, pp.\  281--306.

\bibitem[\protect\citeauthoryear{Fan, Nott, and Sisson}{Fan
  et~al.}{2013}]{fan+ns13}
Fan, Y., D.~J. Nott, and S.~A. Sisson (2013).
\newblock Approximate {B}ayesian computation via regression density estimation.
\newblock {\em Stat\/}~{\em 2\/}(1), 34--48.

\bibitem[\protect\citeauthoryear{Fang, Fang, and Kotz}{Fang
  et~al.}{2002}]{fang+fk02}
Fang, H.-B., K.-T. Fang, and S.~Kotz (2002).
\newblock The meta-elliptical distributions with given marginals.
\newblock {\em Journal of Multivariate Analysis\/}~{\em 82\/}(1), 1 -- 16.

\bibitem[\protect\citeauthoryear{Fasiolo and Wood}{Fasiolo and
  Wood}{2018}]{wood18}
Fasiolo, M. and S.~N. Wood (2018).
\newblock {ABC} in ecological modelling.
\newblock In S.~A. Sisson, Y.~Fan, and M.~A. Beaumont (Eds.), {\em Handbook of
  Approximate Bayesian Computation}. Chapman and Hall/CRC Press.

\bibitem[\protect\citeauthoryear{Fearnhead and Prangle}{Fearnhead and
  Prangle}{2012}]{fearnhead+p12}
Fearnhead, P. and D.~Prangle (2012).
\newblock Constructing summary statistics for approximate {Bayesian}
  computation: semi-automatic approximate {Bayesian} computation.
\newblock {\em Journal of the Royal Statistical Society, Series B\/}~{\em 74},
  419--474.

\bibitem[\protect\citeauthoryear{Feng, Chen, and He}{Feng
  et~al.}{2015}]{fengch2015}
Feng, Y., Y.~Chen, and X.~He (2015).
\newblock Bayesian quantile regression with approximate likelihood.
\newblock {\em Bernoulli\/}~{\em 21\/}(2), 832--580.

\bibitem[\protect\citeauthoryear{Giordani, Mun, Tran, and Kohn}{Giordani
  et~al.}{2013}]{giordani+mtk13}
Giordani, P., X.~Mun, M.-N. Tran, and R.~Kohn (2013).
\newblock Flexible multivariate density estimation with marginal adaptation.
\newblock {\em Journal of Computational and Graphical Statistics\/}~{\em
  22\/}(4), 814--829.

\bibitem[\protect\citeauthoryear{Haario, Saksman, and Tamminen}{Haario
  et~al.}{1999}]{haario+st99}
Haario, H., E.~Saksman, and J.~Tamminen (1999).
\newblock Adaptive proposal distribution for random walk {Metropolis}
  algorithm.
\newblock {\em Computational Statistics\/}~{\em 14}, 375--395.

\bibitem[\protect\citeauthoryear{Isaacs, Altman, Tidmarsh, Valman, and
  Webster}{Isaacs et~al.}{1983}]{Isaac1983}
Isaacs, D., D.~G. Altman, C.~E. Tidmarsh, H.~B. Valman, and A.~D.~B. Webster
  (1983).
\newblock Serum immunoglobulin concentration in preschool children measured by
  laser nephelometry: reference ranges for {IgG}, {IgA}, {IgM}.
\newblock {\em Journal of Clinical Pathology\/}~{\em 36}, 1193--1196.

\bibitem[\protect\citeauthoryear{Koenker}{Koenker}{2005}]{Koenker2005}
Koenker, R. (2005).
\newblock {\em Quantile regression}, Volume~38 of {\em Econometric Society
  Monographs}.
\newblock Cambridge: Cambridge University Press.

\bibitem[\protect\citeauthoryear{Kousathanas, Leuenberger, Helfer, Quinodoz,
  Foll, and Wegmann}{Kousathanas et~al.}{2016}]{kousthanas+lhqfw15}
Kousathanas, A., C.~Leuenberger, J.~Helfer, M.~Quinodoz, M.~Foll, and
  D.~Wegmann (2016).
\newblock Likelihood-free inference in high-dimensional models.
\newblock {\em Genetics\/}~{\em 203}, 893--904.

\bibitem[\protect\citeauthoryear{Leuenberger and Wegmann}{Leuenberger and
  Wegmann}{2010}]{leuenberger+w10}
Leuenberger, C. and D.~Wegmann (2010).
\newblock Bayesian computation and model selection without likelihoods.
\newblock {\em Genetics\/}~{\em 184}, 243--52.

\bibitem[\protect\citeauthoryear{Li, Nott, Fan, and Sisson}{Li
  et~al.}{2017}]{li+nfs15}
Li, J., D.~J. Nott, Y.~Fan, and S.~A. Sisson (2017).
\newblock Extending approximate {B}ayesian computation methods to high
  dimensions via {G}aussian copula.
\newblock {\em Computational Statistics and Data Analysis\/}~{\em 106}, 77--89.

\bibitem[\protect\citeauthoryear{Li and Fearnhead}{Li and
  Fearnhead}{2016}]{li+f16}
Li, W. and P.~Fearnhead (2016).
\newblock Improved convergence of regression adjusted approximate {Bayesian}
  computation.
\newblock {\em arXiv: 1609.07135\/}.

\bibitem[\protect\citeauthoryear{L{\o}land, Huseby, Hjort, and
  Frigessi}{L{\o}land et~al.}{2013}]{loland+hhf13}
L{\o}land, A., R.~B. Huseby, N.~L. Hjort, and A.~Frigessi (2013).
\newblock Statistical corrections of invalid correlation matrices.
\newblock {\em Scandinavian Journal of Statistics\/}~{\em 40\/}(4), 807--824.

\bibitem[\protect\citeauthoryear{Nott, Fan, Marshall, and Sisson}{Nott
  et~al.}{2014}]{nott+fms14}
Nott, D.~J., Y.~Fan, L.~Marshall, and S.~A. Sisson (2014).
\newblock Approximate {B}ayesian computation and {B}ayes linear analysis:
  towards high-dimensional {ABC}.
\newblock {\em Journal of Computational and Graphical Statistics\/}~{\em
  23\/}(1), 65--86.

\bibitem[\protect\citeauthoryear{Nott, Tan, Villani, and Kohn}{Nott
  et~al.}{2012}]{nott+tvk12}
Nott, D.~J., S.~L. Tan, M.~Villani, and R.~Kohn (2012).
\newblock Regression density estimation with variational methods and stochastic
  approximation.
\newblock {\em Journal of Computational and Graphical Statistics\/}~{\em 21},
  797--820.

\bibitem[\protect\citeauthoryear{Ong, Nott, Tran, Sisson, and Drovandi}{Ong
  et~al.}{2017}]{ong+ntsd16}
Ong, V. M.-H., D.~J. Nott, M.-N. Tran, S.~A. Sisson, and C.~C. Drovandi (2017).
\newblock Variational {Bayes} with synthetic likelihood.
\newblock {\em Statistics and Computing\/}, in press.

\bibitem[\protect\citeauthoryear{Peters and Sisson}{Peters and
  Sisson}{2006}]{peters+s06}
Peters, G.~W. and S.~A. Sisson (2006).
\newblock Bayesian inference, {Monte Carlo} sampling and operational risk.
\newblock {\em Journal of Operational Risk\/}~{\em 1}, 27--50.

\bibitem[\protect\citeauthoryear{Rayner and MacGillivray}{Rayner and
  MacGillivray}{2002}]{raynerm02}
Rayner, G. and H.~MacGillivray (2002).
\newblock Weighted quantile-based estimation for a class of transformation
  distributions.
\newblock {\em Computational Statistics \& Data Analysis\/}~{\em 39\/}(4),
  401--433.

\bibitem[\protect\citeauthoryear{Reich, Bondell, and Wang}{Reich
  et~al.}{2010}]{Reich2010}
Reich, B.~J., H.~D. Bondell, and H.~J. Wang (2010).
\newblock Flexible {B}ayesian quantile regression for independent and clustered
  data.
\newblock {\em Biostatistics\/}~{\em 11\/}(2), 337--352.

\bibitem[\protect\citeauthoryear{{Reserve Bank of Australia}}{{Reserve Bank of
  Australia}}{2014}]{rba14}
{Reserve Bank of Australia} (2014).
\newblock Historical data.
  http://www.rba.gov.au/statistics/historical-data.html.

\bibitem[\protect\citeauthoryear{Rodrigues, Nott, and Sisson}{Rodrigues
  et~al.}{2016}]{rodrigues+ns16}
Rodrigues, G.~S., D.~J. Nott, and S.~A. Sisson (2016).
\newblock Functional regression approximate {Bayesian} computation for
  {Gaussian} precess density estimation.
\newblock {\em Computational Statistics and Data Analysis\/}~{\em 103},
  229--241.

\bibitem[\protect\citeauthoryear{Rodrigues and Fan}{Rodrigues and
  Fan}{2017}]{rodrigues16}
Rodrigues, T. and Y.~Fan (2017).
\newblock Regression adjustment for non crossing {B}ayesian quantile
  regression.
\newblock {\em Journal of Computational and Graphical Statistics\/}~{\em 26},
  275--284.

\bibitem[\protect\citeauthoryear{Sisson, Fan, and Beaumont}{Sisson
  et~al.}{2018}]{sisson+fb18}
Sisson, S.~A., Y.~Fan, and M.~A. Beaumont (2018).
\newblock Overview of approximate {Bayesian} computation.
\newblock In S.~A. Sisson, Y.~Fan, and M.~A. Beaumont (Eds.), {\em Handbook of
  Approximate Bayesian Computation}. Chapman and Hall/CRC Press.

\bibitem[\protect\citeauthoryear{Sklar}{Sklar}{1959}]{sklar59}
Sklar, A. (1959).
\newblock Fonctions de repartition a $n$ dimensions et leur marges.
\newblock Publ. Inst. Statist. Univ. Paris 8, 229--231.

\bibitem[\protect\citeauthoryear{Tokdar and Kadane}{Tokdar and
  Kadane}{2012}]{tokdar2012}
Tokdar, S.~T. and J.~B. Kadane (2012).
\newblock Simultaneous linear quantile regression: A semiparametric {B}ayesian
  approach.
\newblock {\em Bayesian Analysis\/}~{\em 7\/}(1), 51--72.

\bibitem[\protect\citeauthoryear{Tran, Nott, and Kohn}{Tran
  et~al.}{2012}]{tran+nk12}
Tran, M.-N., D.~J. Nott, and R.~Kohn (2012).
\newblock Simultaneous variable selection and component selection for
  regression density estimation with mixtures of heteroscedastic experts.
\newblock {\em Electronic Journal of Statistics\/}~{\em 6}, 1170--1199.

\bibitem[\protect\citeauthoryear{Tran, Nott, and Kohn}{Tran
  et~al.}{2017}]{tran+nk16}
Tran, M.-N., D.~J. Nott, and R.~Kohn (2017).
\newblock Variational {Bayes} with intractable likelihood.
\newblock {\em Journal of Computational and Graphical Statistics\/}, in press.

\bibitem[\protect\citeauthoryear{White, Kypraios, and Preston}{White
  et~al.}{2015}]{white+kp15}
White, S.~R., T.~Kypraios, and S.~P. Preston (2015).
\newblock Piecewise approximate {Bayesian} computation: fast inference for
  discretely observed {Markov} models using a factorised posterior
  distribution.
\newblock {\em Statistics and Computing\/}~{\em 25}, 289--301.

\bibitem[\protect\citeauthoryear{Wood}{Wood}{2010}]{wood10}
Wood, S.~N. (2010).
\newblock Statistical inference for noisy nonlinear ecological dynamic systems.
\newblock {\em Nature\/}~{\em 466}, 1102--1104.

\end{thebibliography}
\thispagestyle{empty}
\end{document}